\documentclass[journal,10pt]{IEEEtran}

\usepackage{algorithmic,times}
\usepackage{graphics}
\usepackage{epsfig}
\usepackage{amstext}
\usepackage{amssymb}
\usepackage{amsmath}
\usepackage{subfigure}
\usepackage[font=footnotesize]{caption}
\usepackage{url}
\usepackage{psfrag}
\usepackage{verbatim}
\usepackage{algorithm}
\usepackage{amsthm}
\usepackage{multirow}
\usepackage{color}
\usepackage{soul}
\usepackage{float}

\newcommand\Mark[1]{\textsuperscript#1}

\begin{document}
\title{On the Packet Allocation of \\ Multi-Band Aggregation Wireless Networks}
%Authors Name
\author{\fontsize{11}{13}\selectfont Sanjay Goyal\Mark{1}, Tan Le\Mark{2}, Amith Chincholi\Mark{3}, Tariq Elkourdi\Mark{4}, Alpaslan Demir\Mark{1} \\
\Mark{1} InterDigital Communications, Inc., Melville, NY, USA \\
\Mark{2} Science and Technology Department, Ministry of Information and Communications of Vietnam, Vietnam \\
\Mark{3} Qualcomm Incorporated, USA \\
\Mark{4} New Jersey Institute of Technology, NJ, USA  \\
\small{\texttt{sanjay.goyal}\texttt{@interdigital.com}, \texttt{lebatan}\texttt{@gmail.com}, \texttt{amithc}\texttt{@qti.qualcomm.com}, \texttt{tariq.elkourdi}\texttt{@njit.edu}, \texttt{alpaslan.demir}\texttt{@interdigital.com}
}

\thanks{The final publication is available at Springer via https://link.springer.com/article/10.1007/s11276-017-1486-1.}
}
\maketitle

%%%%%%%%%%%%%%%%%%%%%%%%%%%%%%%%%%%%%%%%%%%
% %%%%%%%%%%%        Abstract           %%%%%%

\begin{abstract}
The use of heterogeneous networks with multiple radio access technologies (RATs) is a system concept that both academia and industry are studying. In such system, integrated use of available multiple RATs is essential to achieve beyond additive throughput and connectivity gains using multi-dimensional diversity. This paper considers an aggregation module called opportunistic multi-MAC aggregation (OMMA). It resides between the IP layer and the air interface protocol stacks, common to all RATs in the device. We present a theoretical framework for such system while considering a special case of multi-RAT systems, i.e., a multi-band wireless LAN (WLAN) system. An optimal packet distribution approach is derived which minimizes the average packet latency (the sum of queueing delay and serving delay) over multiple bands. It supports multiple user terminals with different QoS classes simultaneously. We further propose a packet scheduling algorithm, OMMA Leaky Bucket, which minimizes the packet end-to-end delay, i.e., the sum of average packet latency and average packet reordering delay. We also describe the system architecture of the proposed OMMA system, which is applicable for the general case of the multi-RAT devices. It includes functional description, discovery and association processes, and dynamic RAT update management. We finally present simulation results for a multi-band WLAN system. It shows the performance gains of the proposed OMMA Leaky Bucket scheme in comparison to other existing packet scheduling mechanisms.
\end{abstract}


\section{Introduction}
\label{intro}
The widespread use of multiple radio access technologies (multi-RATs) has attracted many researchers from academia and industry towards the concept of multi-RAT aggregation. With the availability of multiple RATs, the idea of simultaneous use of multiple radios is a viable solution to improve throughput, connectivity, and security~\cite{OMMA1, Intel_multiRAT, ZhangVMH10, KoudouridisYK09, GLL2015,Netgear}. Typical multi-RAT wireless devices support IEEE 802.11 based Wi-Fi RATs like IEEE 802.11n, cellular technologies like UMTS/WCDMA, HSPA, CDMA20001x-EVDO, WiMAX, LTE, GSM, and short range wireless technologies such as Bluetooth. Bandwidth aggregation solutions across multi-RATs could be implemented at different layers such as the application layer, the transport layer, or between the IP and the MAC layers. The aggregation solutions at the transport or the application layers \cite{Han2006, Kelly2005, FRHB11, Key_multipathrouting} may not be very efficient in terms of performance. The lack of instantaneous channel information at these layers makes them inefficient under varying channel conditions. However, the availability of feedback instantaneous channel information from the MAC makes aggregation at a layer between the IP and the MAC \cite{Generic_link_layer_1, Generic_link_layer_2} more promising. 

\begin{figure}
\centering
\includegraphics[width=0.8\linewidth]{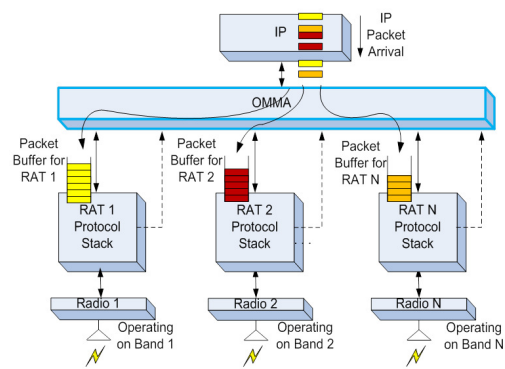}
\caption{Multi-RAT aggregation using OMMA layer}
\label{fig:Overview}
\vspace{-5mm}
\end{figure}

Koudouridis ${\emph{et al.}}$~\cite{Generic_link_layer_1} and Dimou ${\emph{et al.}}$~\cite{Generic_link_layer_2} showed aggregation at a layer between the IP and the MAC. It is called Generic Link Layer (GLL), which is responsible for multi-radio cooperation. It was shown that the GLL can achieve gains in system throughput through efficient utilization of radio resources using multi-radio diversity. On the standardization side, IEEE 802.1 OmniRAN task group is working on a Open Mobile Network Interface (OMNI). It is a common module below the IP layer, enabling simultaneous operation of any IEEE 802 access technology~\cite{HetNet}. In this paper, we consider a similar aggregation module called opportunistic multi-MAC aggregation (OMMA). It resides above the air interface protocol stacks but just below the IP layer and is common to all RATs in the device as shown in Figure~\ref{fig:Overview}. A detailed functional description of the OMMA system will be discussed later in this paper. The details of the key operations required for a multi-RAT system, i.e., RAT capability discovery, association process, and dynamic RAT update management are added to the description. 

Packet or resource scheduling in wireless networks has been widely investigated in the literature for various deployment scenarios, for example~\cite{Zhang2015,Zheng2016, Zhang2014}. In the context of a multi-RAT capable device, multi radio diversity need to be used to extract the maximum gain. To achieve that, a rational packet distribution or resource allocation over all the RATs is essential. Trinh ${\emph{et al.}}$~\cite{Radio_Resource_Switching} proposed a radio resource switching mechanism where resources can be dynamically adjusted between the available RATs. It uses average channel utilization and average packet loss rate to make its decision. To maximize total multi-RAT system capacity, an optimal band selection and power allocation is given in~\cite{choi2010joint}. Chen ${\emph{et al.}}$~\cite{chen2009opportunistic} presented a channel based packet scheduling mechanism to maximize spectral usage of available multiple radios. It collects channel quality feedback over multiple radios simultaneously and schedules multiple transmissions on the available channels accordingly. A utility function based packet scheduling algorithm for GLL, which takes QoS, fairness and spectral efficiency into account is presented by Cui ${\emph{et al.}}$~\cite{cui2009novel}. It provides improvement in packet loss ratio and spectrum efficiency while meeting allowable average packet delay. Koudouridis ${\emph{et al.}}$~\cite{Koudouridis2016} considered a heterogenous small cell networks with multiple radio access (RA) carriers. The solution to the problem of RAs to user association and then selection of RAs for transmission at each user is given in this paper. It was shown that such multi radio transmit diversity can provide over 100\% throughput gain to cell edge users with significant energy efficiency improvements. 

%Su ${\emph{et al.}}$ [22] considered a system with the integration of WLAN and LTE heterogenous networks. An economic model-based network selection algorithm is designed for such system in both under-loaded and over-loaded scenarios.
 A multi-radio resource management for the parallel multi-radio access technology in a cognitive multi-cell is considered by Zhou ${\emph{et al.}}$~\cite{Zhou2016}. For secondary users, authors proposed a proportional fairness based interference management (with primary users) while satisfying the resource constraints caused by multi-radio access. The proposed strategy can achieve the fairness between real-time and the best-effort services for secondary users. Wu ${\emph{et al.}}$~\cite{WuVKHC11} provided an optimal matching between users and RATs with the objective of capacity optimization for the voice and the video communications. Kon ${\emph{et al.}}$~\cite{KonIHHIH12} proposed an autonomous parameter optimization scheme using a machine learning algorithm to maximize throughput of the heterogeneous radio access network (RAN) aggregation system. Interested readers can refer to~\cite{Concurrent_Bandwidth_Aggregation, Survey2012}, and references therein for more multi-RAT bandwidth aggregation schemes. 
 
In this paper, we consider a special case of the multi-RAT systems, i.e., multi-band wireless LAN (WLAN) systems. Our work differs from the existing studies such that we consider a tightly integrated multi-band networks where the objective is to minimize the average packet latency and the reordering delay due to the transmissions over multiple bands. In our previous work~\cite{OMMA1}, we presented an analytical framework for data allocation at the OMMA layer such that average packet latency can be minimized. The previous work covers the case of a single access point (AP) with a single station (STA), and a single type QoS traffic. In this paper, we extend the analysis for a general scenario of multi-STAs, multi-QoS. We investigate the optimal IP packet distribution problem across multiple bands with the objective of minimizing average packet latency, which consists of average queueing delay and average serving delay. In addition, we propose an optimal packet distribution algorithm which also minimizes reordering delay in reference to the minimum average packet latency. We consider that both AP and STAs support simultaneous multi-RAT operations~\cite{Broadcom}.

The key contributions of the paper are summarized as follows:
\begin{itemize}
\item An analytical framework for multi-band WLAN systems with multi-user and multi-QoS scenario to derive the optimal packet allocation ratio over multiple bands.
\item A smart packet allocation algorithm for multi-band systems. It achieves the derived optimal packet allocation ratio and also minimizes the reordering delay at the receiver.
\item Architecture and functional description of the OMMA system for the general case of multi-RAT devices. It includes discovery, association process, and dynamic RAT update management.
\item Simulation results for a multi-band WLAN system comparing the performance of the proposed packet allocation mechanism with the other possible schemes.
\end{itemize}

The rest of the paper is organized as follows. Section~\ref{sec:System Model_PS} describes the system model and the problem statement. The analytical framework for optimal packet allocation over multiple bands is described in Section~\ref{sec:Optimal_Scheduling}. Section~\ref{sec:leakybucket} presents a packet scheduling algorithm for minimizing the re-sequencing delay. The architecture with the functional design of the OMMA system and the flow management at both OMMA sender and receiver are presented in Section~\ref{sec:OMMA_Architecture} and Section~\ref{sec:IP_flow_management}, respectively. Section~\ref{sec:Performance_Evaluation} presents simulation results and analysis. Section~\ref{conclusion} concludes the paper. 

%%%%%%%%%%%%%%%%%%%%%%%%%%%%%%%%%%%%%%%%%%%
% %%%%%%%%%%%        System Description           %%%%%%
\section {System Description}\label{sec:System Model_PS}
% %%%%%%%%%%%        System Model           %%%%%%
\subsection {System Model}\label{sec:System Model}
The wireless system under consideration is a Wi-Fi system. The system consists of an AP, and $N$ number of Wi-Fi STAs. Both AP and STAs have the capability of supporting multiple bands (say $M$ bands), where one MAC layer is associated for each band. The MACs belong to the IEEE 802.11 protocol suite, i.e., 802.11n~\cite{IEEE802.11n}, 802.11ac~\cite{IEEE802.11ac}, etc. As shown in Figure~\ref{fig:Overview}, a common layer called OMMA resides below the IP layer but above the protocol stacks of all bands/RATs. At the AP, a stream of incoming IP packets arrive at the OMMA layer. This incoming IP packet stream is then split by the OMMA layer into $M$ sub-streams each of which is assigned to a corresponding transmit buffer in each band.

\begin{figure}
\centering
\includegraphics[width=0.7\linewidth]{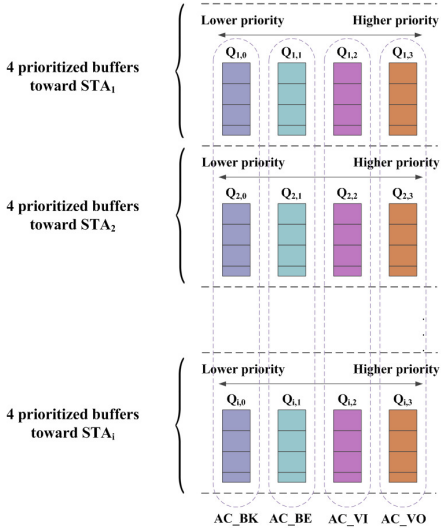}
\caption{MAC queueing mechanism in EDCA mode}
\label{fig:EDCA}
\vspace{-5mm}
\end{figure}

The incoming IP packets at the AP may belong to one of many different IP QoS classes. Each MAC supports enhanced distributed channel access (EDCA)~\cite{IEEE802.11e}, and independently performs mapping of IP QoS classes to 802.11 QoS classes (access categories (ACs)). The packets sent from the OMMA layer to different bands (i.e., sub-streams mentioned above) will be stored in one of four different queues corresponding to four ACs in the MAC~\cite{IEEE802.11e}. Inside each AC queue, there are multiple virtual sub-queues corresponding to each STA that the AP needs to send data to. This queueing structure is modeled as a two dimensional queueing system as shown in Figure \ref{fig:EDCA}.

Let $Q_{i,k}$ denote the queue at the AP which stores packets corresponding to AC $k$ to be sent to STA $i$. Each queue $Q_{i,k}$ is modeled as M/G/1 queue with the following  assumptions:

\begin{itemize}
\item Packets arriving at the MAC layer from the IP layer follow the Poisson process. We acknowledge that the arrival traffic may be bursty, and the Poisson packet arrival assumption will not be valid any more. However, in our analysis, to maintain model tractability, we consider the Poisson packet arrival; which is quite conventional and has been widely used in literature~\cite{data_networks}.
\item Serving time of IP packets follows the general distribution. This is defined as the contention time of the CSMA/CA process plus the transmission time (including retransmissions, if required) for the packet confirmed to be sent out successfully.
\item Packets are served in the order in which they arrive (i.e., first-in-first-out (FIFO)).
\item Serving time of an IP packet is assumed to be identically distributed, mutually independent and independent of the inter-arrival time.
\end{itemize}

The CSMA/CA process of each AC works simultaneously to contend for the channel. All ACs have priorities assigned to them for sending their data to support QoS requirements of different types of traffic. The priority order of ACs is enabled by setting different parameters for the CSMA/CA processes. The AC which wins the contention process will get the channel to send data from its own virtual queues. Packets in the other ACs will remain in their respective queues for this duration. Let us assume that AC $k$ is the winner and currently contends the channel. Inside this AC, there are $N$ independent virtual queues, which store data packets for $N$ different STAs. The AP has different mechanisms to select the STA to be served at this time. If the virtual queue to be scheduled is $Q_{i,k}$, the AP will only send out data corresponding to STA $i$ and AC $k$. This channel access duration may be used to transmit one packet or multiple packets for the queue. Once this transmission completes, the CSMA/CA processes of all the ACs are resumed. After winning the channel, the next AC sends its data. When the channel access is granted to AC $k$ again, the next STA $i+1$ will be served in case they follow the Round-Robin scheduling. In other words, the packets of all the virtual queues corresponding to different STAs need to be served at least once before the first STA is served again. The detail of this process in the EDCA mode can be found in~\cite{next_gen_wlan}.

Due to this queueing system in the EDCA mode, each queue $Q_{i,k}$ can be modeled as M/G/1 queue with vacations. The vacation time of the queue $Q_{i,k}$ is the duration when the AP serves other STAs of the same AC or the other ACs.

\subsection {Problem Statement}\label{sec:Problem_Statement}
In this paper, a packet scheduling problem of IP traffic over multiple bands in the multi-band Wi-Fi devices is considered. As described in Section \ref{sec:System Model}, in our multi-band Wi-Fi system, the main IP stream is split into $M$ sub-streams each of which is assigned to a corresponding transmit buffer in each band.

In a system with single STA, single AP and single IP flow from AP to STA, the challenge is to determine how to optimally distribute packets across bands such that the average end-to-end delay per packet is minimized. Placing all packets in the transmit buffer of the band with the lowest latency may increase the average packet queueing delay. Also, dispersing the packets across all bands randomly may decrease the average packet queueing delay and serving delay. However, it may result in out-of-order reception of packets at the receiver due to differences in link latencies. This can cause longer queueing delays at the receiver to rearrange packets before sending them up to the IP layer. A smart packet assignment strategy to minimize both average end-to-end packet latency and out-of-order packet reception delay, and thus maximizing throughput is essential.

In a system with multiple STAs with multiple IP flows, the MAC layer at the AP sorts the IP flows according to their corresponding ACs and STA addresses. The main IP stream for each STA is split into multiple QoS streams that are queued in separate buffers in the sender MAC layer. Each buffer corresponds to an AC that has a certain priority. However, this set up complicates the traffic shaping. It is because, for any band, the average packet delay not only depends on the queueing delay due to unserved packets in the same buffer, but also on the delay due to buffering and channel access of packets from the other buffers. This is because all packets queued in the MAC from a particular band share the same MAC scheduler and physical layer. Thus, a modified packet assignment strategy compared to the single STA single IP flow case is essential to minimize average end-to-end packet latency.

%%%%%%%%%%%%%%%%%%%%%%%%%%%%%%%%%%%%%%%%%%%
% %%%%%%%%%%%    Optimal Scheduling Scheme at OMMA layer         %%%%%%
\section{Optimal Scheduling Scheme at OMMA Layer}\label{sec:Optimal_Scheduling}
% %%%%%%%%%%%        Terminology and Assumptions        %%%%%%
\subsection{Terminology and Assumptions}\label{sec:terminology}
We use the following terminology shown in Table~\ref{table:single}. Note that the analytical framework developed in this section is for the general case of scheduling data corresponding to AC $k$ from AP to STA $i$. To keep the notation simple, we omit the subscripts $i$ and $k$. The terms defined below is corresponding to band $j \in [1,M]$. Note that,

\begin{table}
\begin{center}
\begin{tabular}{cll}
\hline
&$\lambda_j$ & Average arrival rate of IP packets\\
&$\mu_j$ & Average serving rate\\
&$\rho_j$ & Fraction of arrival rate and service rate, i.e,  $\frac{\lambda_j}{\mu_j}$\\
&$\overline{X_j}$ & Average packet serving time\\
&$\overline{T_j}$ & Total average delay per packet\\
&$\overline{W_j}$ & Average packet queueing delay \\
&$\overline{V_j}$ & Average vacation time \\
\hline
\end{tabular}
\end{center}
\caption{Notations for AC $k$ for STA $i$, i.e., $Q_{i,k}$, corresponding to band $j$ at AP}\label{table:single}
\vspace{-5mm}
\end{table}

\begin{itemize}
\item $\overline{V_j}$ is defined as the average vacation time of the queue $Q_{i,k}$ corresponding to band $j$, i.e., the average time duration that band $j$ stops serving the queue $Q_{i,k}$ and serves queues belonging to other STAs ($\neq i$) or other ACs ($\neq k$).
\item $\overline{V_j^2}$ is the second moment of the average vacation time~$\overline{V_j}$.
\item $\overline{X_j^2}$ is the second moment of average serving time  $\overline{X_j}$.
\end{itemize}

\subsection{M/G/1 Queueing Model with Vacations}\label{sec:qmodel}
The average service time of one packet sent over band $j$ is the inverse of service rate, 
\begin{equation}
\centering
\label{eq:ser_time}
\overline{X_j} = E\{X_j\} = \frac{1}{\mu_j}.
\end{equation}

The second moment of the average service time of packets could be written as:

\begin{equation}
\overline{X_j^2} = E\{X_j^2\}.
\end{equation}

We model queue $Q_{i,k}$ as an M/G/1 queue with vacations. Using the derivation and proof of Pollaczek-Khinchin (P-K) formula for this model as shown in~\cite{data_networks}, the average per packet delay at queue $Q_{i,k} $ corresponding to band $j$ can be written as:

\begin{equation}
\overline{W_j} = \frac{\lambda\overline{X_j^2}}{2(1-\rho_j)} + \frac{\overline{V_j^2}}{2\overline{V_j}}.
\end{equation}

The total delay experienced by one packet in a system is defined as the sum of queueing delay and serving delay. The total average delay for each packet at AP is given by 

\begin{equation}
\label{Delay}
\overline{T_j} = \overline{W_j} + \overline{X_j} = \frac{\lambda_j\overline{X_j^2}}{2(1-\frac{\lambda_j}{\mu_j})} + \frac{\overline{V_j^2}}{2\overline{V_j}} + \overline{X_j}.
\end{equation}

Note that, for each queue $Q_{i,k}$, the parameters $\overline{X_j}$, $\overline{X_j^2}$, $\mu_{j}$, $\overline{V_j}$, $\overline{V_j^2}$ could be measured and fed back by band $j$ to OMMA layer. So, if the arrival rate $\lambda_j$ is known, the average packet delay $\overline{T_j}$ of the AC $k$ toward STA $i$ could be calculated by ($\ref{Delay}$). 

% %%%%%%%%%%%        Optimization Problem Statement       %%%%%%
\subsection{Optimization Problem Statement}\label{sec:statement}
In this section, we formulate an optimization problem to find an optimal scheme at the OMMA layer to distribute the incoming IP traffic corresponding to AC $k$ toward STA $i$ across multiple bands. Assuming there are $M$ bands at the AP to send data to $N$ different STAs. The data sent from queue $Q_{i,k}$ of the AP needs to be scheduled to be sent out on $M$ separate bands. We continue to use subscript $j$ in the following equations to indicate band index $j$.

The total IP packet arrival rate into OMMA, $\lambda$, corresponding to AC $k$ toward STA $i$, is the summation of arrival rates to different bands, i.e.,
\begin{equation}
\lambda = \sum_{j=1}^{M}{\lambda_j}.
\end{equation}

To ensure that the queues do not overflow, we impose the following constraint.
\begin{equation}
\lambda_j < \mu_j \hspace{2mm} \mbox{for} \hspace{1mm} 1\leqslant{j}\leqslant{M}.
\end{equation}

Since $\overline{T_j}$, shown in (\ref{Delay}), is the average packet delay at MAC layer of band $j$ at the AP, the average packet delay over all $M$ bands is the weighted average delay of all $\overline{T_j} \hspace{2mm} \mbox{for} \hspace{1mm} 1\leqslant{j}\leqslant{M}$. The weighting factor for each band $j$ is the ratio of the packet arrival rate on band $j$ to the total packet arrival rate into the OMMA layer, i.e., $\frac{\lambda_j}{\lambda}$. Thus the average packet delay over all $M$ bands is:
\begin{equation}
\label{AverageDelay}
F = \frac{\sum_{j=1}^{M} \left((\frac{\lambda_j\overline{X_j^2}}{2(1-\frac{\lambda_j}{\mu_j})} + \frac{\overline{V_j^2}}{2\overline{V_j}} + \frac{1}{\mu_j})*\lambda_j\right)}{\lambda}.
\end{equation}

Since our objective is to minimize the average packet delay over all $M$ bands, the optimization problem can now be stated as follow:
\begin{equation}
\label{eq:opt1_conv2}
\mbox{Minimize} \hspace{2mm} F = \frac{\sum_{j=1}^{M}{\left(\left(\frac{\lambda_j\overline{X_j^2}}{2(1-\frac{\lambda_j}{\mu_j})} + \frac{\overline{V_j^2}}{2\overline{V_j}} + \frac{1}{\mu_j}\right) * \lambda_j\right)}}{\lambda}
\end{equation}

$\mbox{Subject to:}$
\begin{equation}
\label{eq:constraint1_conv}
\left\{ \begin{array}{ccccl}
\sum_{j=1}^{M}{\lambda_j} &=& \lambda & & \\
\lambda_j & > & 0 & \mbox{for} & 1\leqslant{j}\leqslant{M}\\
-\lambda_j & > & -\mu_j & \mbox{for} &  1\leqslant{j}\leqslant{M}.
\end{array}\right.
\end{equation}

\subsection{The Convexity of the Objective Function}\label{sec:cnvx}
\label{convexity}
In this section, we prove that $F(\lambda_1,\lambda_2,...,\lambda_M)$ is a convex function. We can rewrite $F$ as $F = \sum_{j=1}^{M}{f(\lambda_j)}$, where
\begin{equation}
\label{eq:f_func}
 f(\lambda_j) = \frac{\left(\frac{\lambda_j\overline{X_j^2}}{2(1-\frac{\lambda_j}{\mu_j})} + \frac{\overline{V_j^2}}{2\overline{V_j}} + \frac{1}{\mu_j}\right) * \lambda_j}{\lambda}.
\end{equation}

To prove that $F$ is convex, it is sufficient to prove that $f(\lambda_j)$ is convex. The second derivative of $f(\lambda_j)$ is:
\begin{equation}
\label{eq:derv2}
\frac{\partial^2{f}}{\partial{\lambda_j^2}} = -\frac{\mu_j^3\overline{X_j^2}}
{(\lambda_j - \mu_j)^3 * \lambda}.
\end{equation}

From (\ref{eq:constraint1_conv}), $\lambda_j < \mu_j \hspace{2mm}, \mbox{for} \hspace{1mm} 1\leqslant{j}\leqslant{M}$, which makes (\ref{eq:derv2}) always positive for any value of $\lambda_j$. Since $f(\lambda_j)$ has a positive second derivative, it is strictly convex and so is $F(\lambda_1,\lambda_2,...,\lambda_M)$.

% %%%%%%%%%%%       The Lagrangian optimization method       %%%%%%
\subsection{The Lagrangian Optimization Method}\label{sec:lagrangian}
\label{sec:mindelay}
We use Lagrangian optimization method to solve (\ref{eq:opt1_conv2}). The Lagrangian function for this problem can be written as
\begin{align}
\label{eq:lagrangian_func1}
L(\lambda,\gamma,\beta, \delta) = F(\lambda_1,\lambda_2,...,\lambda_M) -
 \gamma * (\sum_{j=1}^{M}{\lambda_j} - \lambda) -
 \notag\\
 \sum_{j=1}^{M}{\beta_j * (\lambda_j )}
 - \sum_{j=1}^{M}{\delta_j * (- \lambda_j +\mu_j )}.
\hspace{2mm}
\end{align}
where, $\lambda_j \hspace{1mm}$ for $1\leqslant{j}\leqslant{M}$ are the unknown variables, and $\gamma,\beta_j, \delta_j \hspace{1mm} \mbox{for} \hspace{1mm} 1\leqslant{j}\leqslant{M}$ are the Lagrangian multipliers. Since the objective function $F$ is a convex function, there must exist an optimal solution set $(\lambda_j^*,\gamma^*,\beta_j^*, \delta_j^*)$, for $1\leqslant{j}\leqslant{M}$. Using the Karush-Kuhn-Tucker conditions~\cite{opt_theo}, the optimal solution has to satisfy the following set of equations:
\begin{equation}
\label{eq:lagrangian_opt}
\left\{ \begin{array}{lclll}
\frac{\partial{F}}{\partial{\lambda_j}} - \gamma - \beta_j + \delta_j & = & 0 &\mbox{for} \hspace{1mm} 1\leqslant{j}\leqslant{M}&\hspace{0mm} \mbox{(a)}\\
\gamma*(\sum_{j=1}^{M}{\lambda_j} - \lambda) & = & 0 & &\hspace{0mm} \mbox{(b)}\\
\beta_j * \lambda_j &=& 0 & \mbox{for} \hspace{1mm} 1\leqslant{j}\leqslant{M}&\hspace{0mm} \mbox{(c)}\\
\delta_j * (\mu_j - \lambda_j) &=& 0 & \mbox{for} \hspace{1mm} 1\leqslant{j}\leqslant{M}&\hspace{0mm} \mbox{(d)}
\end{array}\right.
\end{equation}

In the above set of equations, there are total $3M+1$ number of equalities. From which, there are total $3M+1$ variables need to be derived, which include $\gamma^*$, $\boldsymbol{\Lambda^*} = (\lambda_1^*,\lambda_2^*,...,\lambda_M^*)$, $\boldsymbol{\beta^*} = (\beta_1^*,\beta_2^*,...,\beta_M^*)$, and $\boldsymbol{\delta^*} = (\delta_1^*,\delta_2^*,...,\delta_M^*)$. 

To solve it, please note in (\ref{eq:lagrangian_opt})(c), since $\lambda_j > 0 $, we have $\beta_j = 0$ for $1\leqslant{j}\leqslant{M}$. In (\ref{eq:lagrangian_opt})(d), since $\mu_j > \lambda_j $, we have $\delta_j = 0$ for $1\leqslant{j}\leqslant{M}$. It also implies that the arrival rate is smaller than the equivalent service rate, so the total incoming traffic will always be served optimally such that $\sum_{j=1}^{M}{\lambda_j} = \lambda$. Using these results in (\ref{eq:lagrangian_opt}), we get,
\begin{equation}
\label{eq:lagrangian_sol1}
\left\{ \begin{array}{lclll}
\frac{\partial{F}}{\partial{\lambda_j}} - \gamma & = & 0 & \mbox{for} \hspace{1mm} 1\leqslant{j}\leqslant{M}&\hspace{2mm} \mbox{(a)}\\
\sum_{j=1}^{M}{\lambda_j} - \lambda & = & 0 & &\hspace{2mm} \mbox{(b)}\\
\beta_j & = & 0 &\mbox{for} \hspace{1mm} 1\leqslant{j}\leqslant{M}&\hspace{2mm} \mbox{(c)}\\
\delta_j & = & 0 &\mbox{for} \hspace{1mm} 1\leqslant{j}\leqslant{M}&\hspace{2mm} \mbox{(d)}
\end{array}\right.
\end{equation}

Moreover, (\ref{eq:lagrangian_sol1})(a) can be written as
\begin{align}
\label{eq:lagrangian_sol6}
(\mu_j \overline{V_j^2} + 2 \overline{V_j}-\mu_j^2 \overline{V_j} \overline{X_j^2} - 2\mu_j\overline{V_j}\lambda\gamma)*\lambda_j^2 +
\notag\\
(-2\mu_j^2\overline{V_j^2}+2\mu_j^3\overline{V_j}\overline{X_j^2}-4\mu_j\overline{V_j}+ 4\mu_j^2\overline{V_j}\lambda\gamma)*\lambda_j+
\notag\\
(\mu_j^3\overline{V_j^2}+2\mu_j^2\overline{V_j} - 2\mu_j^3\overline{V_j}\lambda\gamma)= 0,
\end{align}
which gives,
\begin{equation}
%\small
\label{eq:lagrangian_sol7}
\lambda_j^*=
\left(  \mu_j \pm \frac{\mu_j^2\sqrt{\overline{V_j} \overline{X_j^2}}}{\sqrt{\mu_j^2\overline{V_j} \overline{X_j^2}-\mu_j\overline{V_j^2}+(2 \lambda \gamma \mu_j -2 )\overline{V_j}}}\right). \
\end{equation}

Since the objective function is strictly convex, it has a unique non-negative globally optimal solution $\lambda_j, 1\leqslant{j}\leqslant{M}$. Note that in (\ref{eq:lagrangian_sol7}), the solution $\lambda_j$ still contains the unknown variable $\gamma$. Using (\ref{eq:lagrangian_opt})(b) and (\ref{eq:lagrangian_sol7}), 
\begin{equation}
%\small
\label{eq:lagrangian_sol10}
\sum_{j=1}^{M}
\left(  \mu_j \pm \frac{\mu_j^2\sqrt{\overline{V_j} \overline{X_j^2}}}{\sqrt{\mu_j^2\overline{V_j} \overline{X_j^2}-\mu_j\overline{V_j^2}+(2 \lambda \gamma \mu_j -2 )\overline{V_j}}}\right),\
=\lambda
\end{equation}
which is equivalent to:
\begin{equation}
%\small
\label{eq:lagrangian_sol13}
\sum_{j=1}^{M}
\left( \mp \frac{1}{\sqrt{\frac{2\lambda}{\mu_j^3}\gamma  + \left( \frac{1}{\mu_j^2} - \frac{\overline{V_j^2}}{\mu_j^3\overline{V_j}\overline{X_j^2}} - \frac{2}{\mu_j^4\overline{X_j^2}}\right)}}
\right)
=  \sum_{j=1}^{M} \mu_j  - \lambda.
\end{equation}

We make an assumption that $2\lambda \gg \mu_j$ for $1\leqslant{j}\leqslant{M}$, which gives, $\Big( \frac{1}{\mu_j^2} - \frac{\overline{V_j^2}}{\mu_j^3\overline{V_j}\overline{X_j^2}} - \frac{2}{\mu_j^4\overline{X_j^2}}\Big)$ $\ll \Big(\frac{2\lambda}{\mu_j^3} \Big)$. Thus the term $\Big( \frac{1}{\mu_j^2} - \frac{\overline{V_j^2}}{\mu_j^3\overline{V_j}\overline{X_j^2}} - \frac{2}{\mu_j^4\overline{X_j^2}}\Big)$ in (\ref{eq:lagrangian_sol13}) becomes negligible. The above assumption is made to model the situations when congestion happens and packets reach the receiver in an out of order fashion. It is in-fact one of the main problems in the multi-RAT systems such as OMMA, where all the RATs are used simultaneously. One of the main contributions of this paper is to provide a mechanism to handle this problem, which is given in Section~\ref{sec:leakybucket}.

Based on the above assumption, (\ref{eq:lagrangian_sol13}) can be approximated as:
\begin{equation}
\label{eq:lagrangian_sol14}
\sum_{j=1}^{M}
\left( \mp \frac{1}{\sqrt{\frac{2\lambda}{\mu_j^3}\gamma }}
\right)
\approx  \sum_{j=1}^{M} \mu_j  - \lambda,
\end{equation}
which gives,
\begin{equation}
\label{eq:lagrangian_sol17}
\gamma
\approx  \frac {\left(\sum_{j=1}^{M}  \mp \mu_j^{\frac{3}{2}}\right)^2}  {2\lambda \left( \sum_{j=1}^{M} \mu_j  - \lambda\right)^2}.
\end{equation}

Using the above value of $\gamma$ in (\ref{eq:lagrangian_sol7}), we get the values of $\lambda_i^* \hspace{1mm} \mbox{for} \hspace{1mm} 1\leqslant{j}\leqslant{M}$ as

\begin{equation}
%\small
\begin{split}
\label{eq:lagrangian_sol18}
&\lambda_j^*= \\
&\mu_j \pm \frac{\mu_j^2\sqrt{\overline{V_j} \overline{X_j^2}}}{\sqrt{\mu_j^2\overline{V_j} \overline{X_j^2}-\mu_j\overline{V_j^2}+\left(\frac {\left(\sum_{j=1}^{M}  \mp \mu_j^{\frac{3}{2}}\right)^2}  {\left( \sum_{j=1}^{M} \mu_j  - \lambda\right)^2} * \mu_j -2 \right)*\overline{V_j}}}.
\end{split}
\end{equation}

Note that in the above equation, we have up to $2^M$ different solutions due to $\mp$ sign. Each candidate solution set  $\boldsymbol{\Lambda} = (\lambda_1,\lambda_2,...,\lambda_M)$ need to be checked to satisfy the constraints given in (\ref{eq:constraint1_conv}). As soon as a local minimum solution is found, this process can be terminated. It is because of the property of the convex problem, where a local minimum will also be the global minimum solution $\boldsymbol{\Lambda^*}$. Note that when $2\lambda \gg \mu_j $ is not true, then the values of $\lambda_j^*$ can be found by solving (\ref{eq:lagrangian_sol7}) and (\ref{eq:lagrangian_sol13}).

%%%%%%%%%%%%%%%%%%%%%%%%%%%%%%%%%%%%%%%%%%%
% %%%%%%%%%%%    Packet flow control         %%%%%%
\section{Packet Flow Control} \label{sec:leakybucket}
In a multi-RAT system such as OMMA, which performs aggregation on a packet basis, re-sequencing delay is a critical factor that needs to be addressed. Re-sequencing delay for a packet can be defined as as the time the packet needs to wait at the receiver's OMMA layer for all of the earlier packets in sequence to be successfully received. It happens when data packets are received out of order due to packets traversing over the multiple links, each with different packet latency. The OMMA layer may incur re-sequencing delays while reordering the packets before sending them to the IP layer. The re-sequencing problem has a severe impact on both UDP and TCP applications~\cite{QoS_reorder_1,QoS_reorder_2}. For example, the QoS of real-time UDP applications like voice over IP or live video streaming could suffer because out of order packets will be counted as lost packets and will be ignored at the receiver side~\cite{QoS_reorder_2}. For TCP, it is even more serious because out of order packets could generate duplicate ACK issues, which triggers an unnecessary congestion control mechanism that reduces the effective throughput.

At the transmitter side, packets of the main stream are split into multiple sub streams for transmission over different links which may possibly have different latencies. An optimal packet assignment strategy at the OMMA transmitter is necessary to minimize the packet reordering delay at the receiver. In order to achieve that, we propose an algorithm, OMMA Leaky Bucket, which is described in Algorithm~\ref{alg:LeakyBucket}.

\begin{algorithm}
\caption{\label{alg:LeakyBucket} OMMA Leaky Bucket Algorithm}
\label{alg1}
\begin{algorithmic}[1]
\FOR {$j=1$ to $M$}
\STATE  $T_j \Leftarrow 0$
\ENDFOR
\WHILE {Unscheduled packets set $\neq \emptyset$}
\FOR {$j=1$ to $M$}
\STATE  Update $\mu_j$ from Meta Data Feedback of MAC layer
\STATE  Update $\lambda_j^*$ by (\ref{eq:lagrangian_sol18})
\STATE  $R_j \Leftarrow \frac{\lambda_j^*}{\mu_j}$
\ENDFOR
\WHILE {$\forall~T_j < 1, 1\leqslant{j}\leqslant{M}$}
\FOR {$j=1$ to $M$}
\STATE  $T_j \Leftarrow T_j + R_j$
\ENDFOR
\ENDWHILE
\STATE find $i$ with $T_i = $max $\mathcal{T} \{T_1,T_2,...,T_M\}$
\STATE send current packet on band $i$
\STATE $T_i \Leftarrow T_i-1$
\ENDWHILE
\end{algorithmic}
\end{algorithm}
\setlength{\textfloatsep}{20pt}

In this algorithm we maintain $M$ token variables $T_j$ for $1\leqslant{j}\leqslant{M}$, one for each band. Initially, each band is assigned zero token (Lines 1-3). Further, token for each band $j$ is incremented iteratively by $\frac{\lambda_j^*}{\mu_j}$ (Lines 5-14) until at least one of the tokens exceeds 1 (Line 10). Here, $\lambda_j^*$ is the optimal rate for band $j$ calculated by the minimum delay algorithm presented in Section \ref{sec:Optimal_Scheduling}. The band corresponding to the token which exceeds 1 is chosen to send the next incoming unscheduled packet at OMMA (Lines 15-16). After that the token for the selected band is decremented by 1 (Line 17) and the process of incrementing tokens is continued as before (Line 4). This algorithm runs ``ahead'' of every packet arriving at OMMA, i.e., OMMA always knows which packet ID will be scheduled on which band. The $\lambda_j^*$ values used in this algorithm are derived to achieve minimum average delay per packet. Therefore, this algorithm ensures that the band chosen to send each packet is such that, (i) the packet experiences the minimum delay, and (ii) it also arrives in the correct order at the receiver with respect to its preceding and succeeding packets. It basically schedules the packets such that packets coming earlier are sent on the band with lower end-to-end latency, while the later packets are on the higher latency bands.

We call Algorithm~\ref{alg:LeakyBucket} as OMMA Leaky Bucket algorithm since it works analogous to the Leaky Bucket algorithm~\cite{data_networks}. In Algorithm~\ref{alg:LeakyBucket}, the token variables represent bucket to control the rate of scheduling; however, the size of each bucket is not constant. The token increases with the variable rate equals to the ratio of optimal rate allocation and the service rate for the corresponding band. The bucket that leaks by sending the packet is the bucket with the largest size at each particular time, given that its token exceeds 1.

%%%%%%%%%%%%%%%%%%%%%%%%%%%%%%%%%%%%%%%%%%%
% %%%%%%%%%%%    OMMA Architecture         %%%%%%
\begin{figure}
\centering
\includegraphics[width=0.6\linewidth]{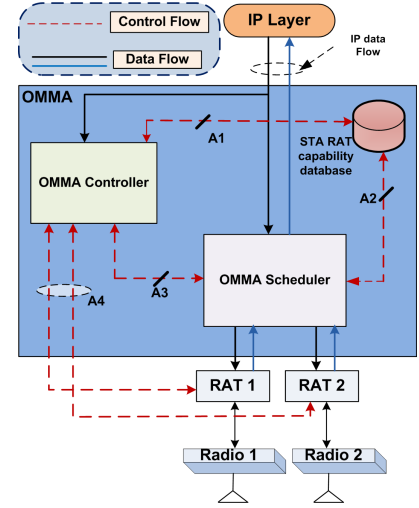}
\caption{Block Diagram of the OMMA Layer}
\label{fig:OMMA_Block_Diagram}
\end{figure}

\section{OMMA Architecture}\label{sec:OMMA_Architecture}
This section describes the architecture of the OMMA layer and its main functional modules. This section also provides the details of some key operations performed at OMMA which are required to support multi-RAT aggregation. Please note that the details presented in this section as well as the next section are applicable for the general case of multi-RAT systems.

\subsection{Functional Description}
A high level architectural view of OMMA is shown in Figure \ref{fig:OMMA_Block_Diagram}. It includes all functional blocks of the OMMA layer including the main interfaces for control signaling and data signaling. The OMMA layer consists of the following main functional blocks.

\subsubsection{STA RAT Capability Database}
At the AP, STA RAT capability database is used to store RAT capability information (i.e., list of all common RATs) for each of its associated STAs. Moreover, because of poor link quality due to interference or mobility, a subset of RATs may be unavailable for a STA. This database also stores a list of available RATs at a given time for each associated STA. This information of RAT capability and available RATs is updated by the OMMA controller as described later.

\subsubsection{OMMA Controller}
The OMMA controller is responsible for updating the STA RAT capability database either in case of newly associated STAs or change in availability of RATs for already associated STAs. The OMMA controller also receives feedback metrics $\mu_j$, $\overline{V_j}$, $\overline{V_j^2}$ and $\overline{X_j^2}$ from each RAT $j$ ($1\leqslant j\leqslant M$) corresponding to AC $k$ from AP to STA $i$ for that RAT. It then classifies those metrics based on the STA ID $i$ and QoS class $k$, and sends them to OMMA schedulers of the corresponding STA. It also calculates the arrival rate $\lambda_j$ (corresponding to AC $k$ from AP to STA $i$) of incoming IP packets and provides this information to the OMMA scheduler. This is one of the parameters required to calculate the optimal split of IP packets across multiple RATs. Moreover, the OMMA controller provides system parameters (e.g. number of RATs, type of RATs, matched set of RATs with STAs to be associated) during discovery and association process as described later.

\subsubsection{OMMA Scheduler}
The OMMA layer maintains a separate OMMA scheduler module corresponding to each associated STA and each QoS class supported by the system. The OMMA layer also maintains an IP packet STA classifier module and also a STA QoS Classifier module to read the IP packet header and send it to the corresponding OMMA scheduler module for further processing. The OMMA scheduler communicates with the STA RAT capability database to extract the list of available RATs for a STA. It selects RATs based on the feedback metrics provided by the OMMA controller and the list of available RATs for that STA provided by STA RAT capability database. On the transmitter side, it distributes packets across selected RATs based on a given packet assignment scheme. On the receiver side, the OMMA scheduler is responsible for aggregating packets received from RATs and sending them to the IP layer.

\subsection{Key operations at OMMA}
This section describes some key operations that are important to enable communication between multi-RAT devices.

\subsubsection{RAT Capability Discovery}
Each multi-RAT device can have a different RAT capability (i.e., set of supported RATs). This generates the need for a discovery and association process in which a device (STA/AP) can advertise its RAT capability parameters (i.e., number of RATs, type of RATs, etc.) to other devices. This way, a STA and an AP can associate with each other on the set of RATs common to them. 

The AP can  advertise its  RAT capabilities either in  the beacon (in the passive scanning mode) or in the Probe Response (in the active scanning mode) which is generated in response to a Probe Request from the STA. The beacon is sent on all the available RATs at the AP while the Probe Response is sent on the same RAT on which the Probe Request was received. The STA, which receives AP's RAT capabilities, selects the set of RATs common to itself and the AP with the help of the OMMA controller. The STA signals the set of common RATs in the Association Request message sent to the AP on every common RAT. The AP stores the information of RAT capabilities of the STA in its STA RAT capability database.

\subsubsection{OMMA Mode Selection}
This procedure is required to decide the mode of operation of the OMMA scheduler at both sender and receiver. The modes of operation could either be based on a pre-defined set of policies for every IP flow, or could be based on feedback parameters received by OMMA from each RAT. Some examples of OMMA modes are described in Section \ref{sec:Performance_Evaluation} (referred to as packet scheduling schemes). At an AP, the OMMA controller makes the mode selection decision and signals this decision to OMMA scheduler. The OMMA scheduler enables/disables packet transmission on certain RATs based on the mode decision. Furthermore, the OMMA controller at an AP sends mode decisions to the OMMA receiver at a STA using one of the available RATs for that STA. At a STA, mode information received in the beacon is signaled to the OMMA controller, which in turn configures the OMMA scheduler accordingly.

\subsubsection{RAT Availability Update Management}
Since the wireless link on each RAT may have variable link quality parameters such as packet loss rate, jitter due to factors such as interference, mobility. Thus some of the RATs common between the STAs may be usable while the others may not be usable. Thus the AP transmitting data to a STA may not be aware of which RATs are usable at any given time.

A procedure for dynamic management of RAT availability for every STA-AP pair addresses this problem. The AP sends beacons on all its RATs periodically. The STA reads the beacon information on all the RATs common to itself and the AP. If the STA is able to read the beacon information successfully on any RAT, it identifies that RAT as being available and assigns a value `1' to that RAT. If the STA is unable to read the beacon information successfully on any RAT, it identifies that RAT as being unavailable and assigns a value `0' to that RAT. Thus the OMMA controller at the STA generates a binary vector of length equal to the number of RATs common to itself and the AP. Each bit in the binary vector indicates whether a RAT is available or not. Note that the different link quality metrics (e.g., RSSI, SINR, BER, etc.) can be used during the process of beacon identification. However, the accuracy of the several link quality metrics varies under different traffic conditions and transmission rates~\cite{vlavianos2008}. It was shown in~\cite{vlavianos2008}that none of the existing link quality metric is sufficient to accurately characterize the quality of link in different conditions. Thus a combination of different metrics could yield more accurate representation of the link quality.

The STA periodically sends this RAT availability binary vector to the AP using one of the common RATs. This information is sent from the RAT to the OMMA controller which in turn stores it in the STA RAT capability database. In this way, the RAT Availability information of any STA-AP pair is periodically refreshed.

%%%%%%%%%%%%%%%%%%%%%%%%%%%%%%%%%%%%%%%%%%%
% %%%%%%%%%%%    IP Flow Management at OMMA       %%%%%%
\section{IP Flow Management at OMMA}\label{sec:IP_flow_management}
This section describes the flow management at both the OMMA sender and receiver for incoming IP packets. The procedure of RAT selection for incoming IP packets at the OMMA sender is described. We also describe the operations at OMMA receiver required to send IP packets (i.e., received from multiple RATs) to IP layer. In this work all IP packets are taken of \emph{single QoS class}.

\begin{figure}
\centering
\includegraphics[width=0.7\linewidth]{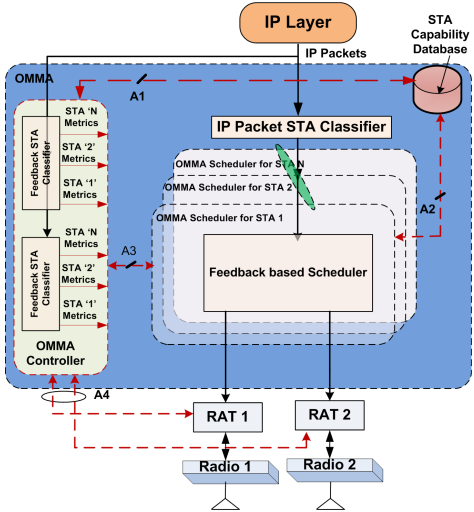}
\caption{OMMA Sender}
\label{fig:OMMA_Sender}
\end{figure}

\subsection{OMMA Sender Operation}
A high level view of the OMMA sender is shown in Figure~\ref{fig:OMMA_Sender}. The OMMA sender takes the decision of RAT selection to send the incoming IP packets. The procedure for routing of incoming packets to a subset of RATs is described below.

\begin{enumerate}
  \item At OMMA, the incoming packet is delivered to both the \emph{IP Packet STA Classifier} and the OMMA controller,
  \item IP packet STA classifier sends packet to OMMA scheduler corresponding to its destined STA,
  \item Scheduler makes the decision on RAT/RATs selection by using RAT availability provided by the STA RAT capability database and feedback metrics (arrival rate, serving rate, and average packet delay) provided from the OMMA controller,
      \begin{itemize}
        \item It selects all the RATs which fulfill the minimum requirement of QoS class of the incoming packet.
        \item In case of starting phase, when there is no feedback available, it chooses all the available RATs for that STA.
      \end{itemize}
  \item The scheduler distributes all the packets on the selected RATs based on the algorithms described in Section~\ref{sec:Performance_Evaluation}.
%  \item RAT switch happens when one of the RATs of multiple selected RATs is not able to fulfill the requirement of given QoS class. In this situation, it chooses randomly a RAT from the set of other available RATs that are not currently chosen for that STA.
\end{enumerate}

\begin{figure}
\centering
\includegraphics[width=0.7\linewidth]{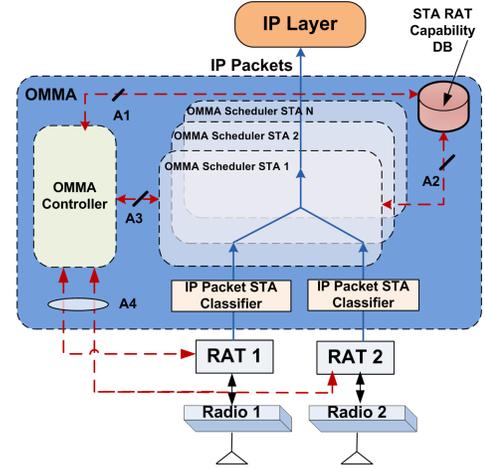}
\caption{OMMA Receiver}
\label{fig:OMMA_Receiver}
\end{figure}

\subsection{OMMA Receiver Operation}
A high level view of OMMA receiver is shown in Figure \ref{fig:OMMA_Receiver}. At the OMMA receiver, OMMA maintains a separate IP packet STA classifier for each RAT. Each IP packet STA classifier reads the packet header and sends it to the OMMA scheduler corresponding to that STA. The OMMA scheduler aggregates the data packets received from multiple RATs and sends them to IP layer.

%%%%%%%%%%%%%%%%%%%%%%%%%%%%%%%%%%%%%%%%%%%
% %%%%%%%%%%%    Performance Evaluation       %%%%%%
\section{Performance Evaluation}\label{sec:Performance_Evaluation}
% %%%%%%%%%%%    Single AP - Single Station       %%%%%%
We use OPNET V16.0 simulator to simulate two different scenarios. In the first scenario, we consider the transmission from a AP to a STA over two WLAN bands. In which one band is an IEEE 802.11n band operating on 2.4 GHz ISM band over single 20 MHz channel. The other band is a proprietary modified IEEE 802.11n band operating on the television whitespace (TVWS) band. In the second scenario, we simulate a single AP communicating with two STAs simultaneously over two WLAN bands described above. Further details of each scenario is given in the next subsections.
\subsection{Single AP - Single Station}\label{subsec:singleap_singleST}

\begin{figure*}[t]
\centering
\includegraphics[width=5in]{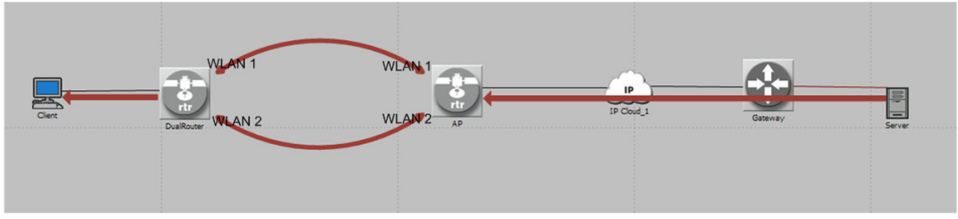}
\caption{Simulation setup for \textit{Single AP - Single Station} scenario}
\label{fig:omma_system}
\end{figure*}

In this section, we present a simulation scenario with single server sending data to a single STA over an IP backhaul network and through the AP as shown in Figure \ref{fig:omma_system}. The downlink data sent from the server is set as best effort traffic (all data belongs to a single QoS class). The ON-OFF traffic model~\cite{schwartz1996broadband} is used to generate the traffic, where the transition time follows an exponential distribution and the arrival IP packets follow the Poisson process. It complies with our analysis given in Section~\ref{sec:Optimal_Scheduling}. Both AP and STA are capable of supporting two WLAN bands, i.e., ISM and TVWS band. The TVWS band is capable of aggregating four TVWS channels (5 MHz per channel) at the MAC layer. The OMMA layer resides on top of the protocol stacks of two bands and below the IP layer at the AP and the client. It is responsible for distributing IP packets across bands at the sending node and collecting them over the two bands at the receiving node. The serving band meta-data is measured at each band and is fed back to the OMMA layer. When downlink IP traffic reaches the AP, the OMMA layer at the AP either sends all traffic over one band or distributes the traffic across two bands. The server sends successive multiple files to the STA by setting up a TCP connection per file transfer. Once a file is completely downloaded, the TCP connection corresponding to it is terminated and a new TCP connection is setup for the next file. We evaluate the performance of several packet allocation schemes at OMMA in terms of TCP throughput, packet latency, and number of retransmissions. The different packet scheduling schemes used at the OMMA layer are:

\begin{itemize}
\item {ISM Band Only}: The AP sends all the data to STA over the MAC operating on ISM band only. The MAC operating on TVWS band is disabled.
\item {TVWS Band Only}: The AP sends all the data to STA over the MAC operating on TVWS band only. The MAC on ISM band is disabled.
\item {50\% - 50\% Traffic Split}: Half of the incoming IP packets at the AP are sent over the MAC operating on ISM band while the other half of the packets are sent over the MAC operating on TVWS band. Packets are assigned to bands sequentially as they arrive at the AP (regardless of the packet ID).
\item {Load Balancing}: Incoming IP packets are assigned to the two bands with respect to the serving rates of bands. Packets are assigned to bands sequentially as they arrive at the AP (regardless of the packet ID).
\item {Band Switching Per TCP Flow}: Packets of a single TCP flow assigned to the same band. One of the bands is selected at any given time using the Round-Robin scheme.
\item {Minimum Delay}: Incoming IP packets are assigned to the two bands with respect to optimal packet distribution scheme determined by the average delay per packet minimization as presented in Section \ref{sec:Optimal_Scheduling}. Packets are assigned to bands sequentially as they arrive at the AP (regardless of the packet ID) but still maintain the optimal packet distribution ratio.
\item {Leaky Bucket}: Incoming IP packets are assigned to the two bands based on the optimal packet distribution scheme determined by the average delay per packet minimization as presented in Section \ref{sec:Optimal_Scheduling}. However, packets are smartly assigned to bands using the OMMA Leaky Bucket technique to minimize both per-packet-delay and out-of-order packet reception as presented in Section \ref{sec:leakybucket}.
\end{itemize}

\begin{table}[h]
\begin{center}
\begin{tabular}{|c|c|c|}
	\hline
No &  Scheme   &   Offered Load (Mbps) \\
	\hline
1   & ISM Band Only& 10 \\
2   & TVWS Band Only& 17.5 \\
3   & 50\% - 50\% Traffic Split & 18 \\
4   & Band Switching Per TCP Flow & 16 \\
5   & Load Balancing & 27 \\
6   & Minimum Delay & 27 \\
7  & Leaky Bucket & 27 \\
	\hline
\end{tabular}
\end{center}
\caption{\label{tbl:offerload} Offered load for each scheme in \textit{Single AP - Single STA} scenario}
\end{table}

The noise power is set independently for both the bands. It is set such that the average SNR level is 20 dB for the TVWS band that consists of four TVWS channels with 5 MHz each, and 10 dB for the ISM band with a single 20 MHz channel. Every 30s, the server sends a new file to the client, which creates a new TCP connection. The file size determines the offered load from application to transport layer. In our simulations, we adjust the file size such that the maximum sustainable offered load would be achieved for each scheduling scheme. The value of the offered load for each scheme that we simulated is shown in Table \ref{tbl:offerload}.

\subsubsection{Small Round Trip Time}
The packet latency within the IP cloud module shown in Figure \ref{fig:omma_system} is set to be negligible. It leads to an average end to end round trip time (RTT) as small as 50 ms with aggregation schemes and 100 ms with single band schemes. In addition, the average MAC latency difference between two bands is around 15 ms. It is equivalent to around 15\% of the RTT in the single band schemes and 30\% of the RTT for the aggregation schemes with multiple bands.

\begin{figure}
\centering
\includegraphics[width=0.8\linewidth]{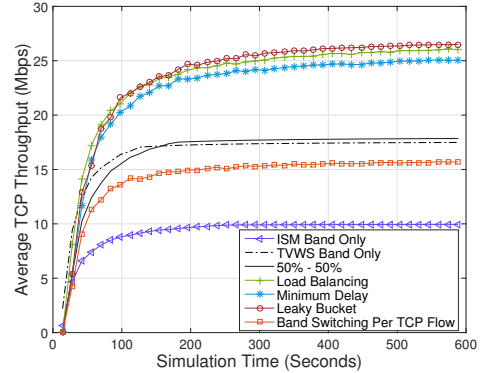}
\caption{Average TCP throughput with small RTT scenario in \textit{Single AP - Single STA} case}
\label{fig:Scenario 1 throughput}
\end{figure}

\begin{figure}[h]
\centering
\includegraphics[width=0.8\linewidth]{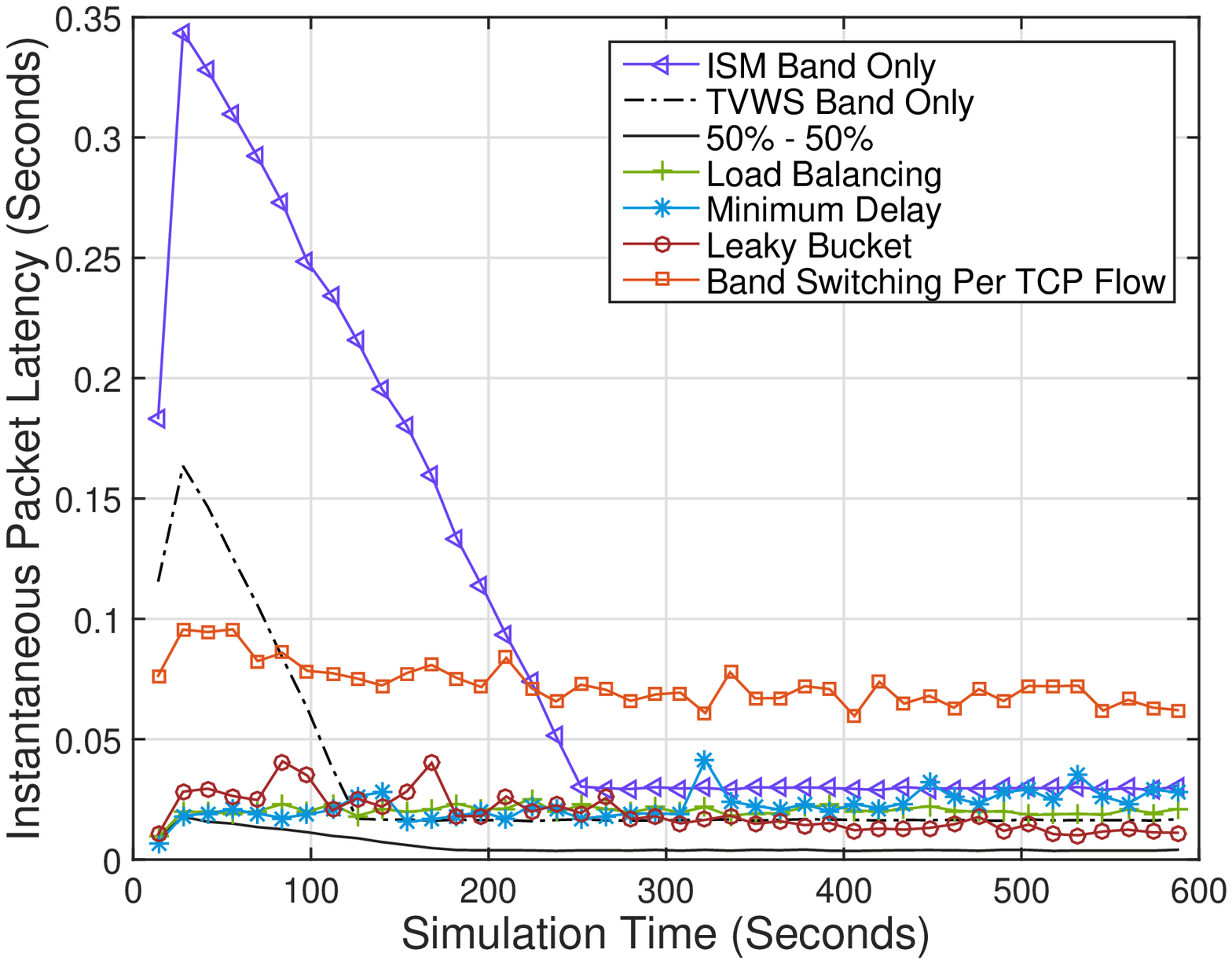}
\caption{Instantaneous packet latency with small RTT scenario in \textit{Single AP - Single STA} case}
\label{fig:Scenario 1 latency}
\vspace{-5mm}
\end{figure}

\begin{figure}[htb]
\centering
\includegraphics[width=0.8\linewidth]{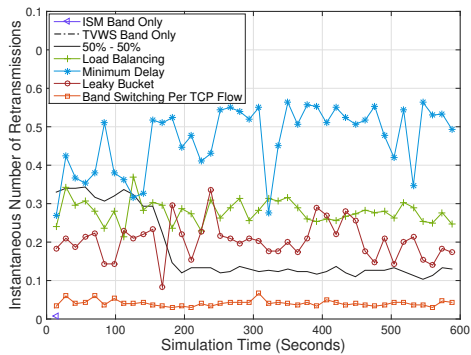}
\caption{Average instantaneous number of retransmissions with small RTT scenario in \textit{Single AP - Single STA} case}
\label{fig:Scenario 1 retrans}
\end{figure}

The performance comparison for scheduling schemes with small RTT scenario is captured in Figures~\ref{fig:Scenario 1 throughput}, \ref{fig:Scenario 1 latency} and \ref{fig:Scenario 1 retrans}. These figures show the average throughput, instantaneous packet latency, and the average number of instantaneous retransmissions, respectively. The throughput performance of the different packet allocation schemes shown in Figure \ref{fig:Scenario 1 throughput} is in consonance with the offered load shown in Table \ref{tbl:offerload}. The aggregation schemes such as \textit{Load Balancing}, \textit{Minimum Delay}, and \textit{Leaky Bucket} are able to maintain higher offered load so as to achieve highest throughput due to the adaptive scheduling over multiple bands using the channel metric feedback on each band. Moreover, the proposed \textit{Leaky Bucket} scheme provides the highest throughput compared to the \textit{Minimum Delay} and the \textit{Load Balancing} schemes while maintaining the same highest offered load. It is due to the smart scheduling per packet, that helps in reducing the reordering delay and achieving the less number of duplicate ACKs. Therefore, among all the aggregation schemes with the highest offered load, the \textit{Leaky Bucket} scheme is able to maintain the lowest number of retransmissions as shown in Figure~\ref{fig:Scenario 1 retrans}, which also makes it best in terms of latency reduction as shown in Figure~\ref{fig:Scenario 1 latency}.

The single band schemes suffer high packet latency at the beginning of the simulation but later converge to the expected level as shown in Figure~\ref{fig:Scenario 1 latency}. In the \textit{ISM Band Only} scheme, the packets experience higher packet latency with longer convergence time. The TCP flow control needs more time to absorb the network load on the low SNR at ISM band branch. This is also the reason for lower throughput performance of the \textit{ISM Band Only} compared to the \textit{TVWS Band only} as shown in Figure~\ref{fig:Scenario 1 throughput}. It is also interesting to see that the \textit{50\% - 50\% Traffic Split} has almost the same throughput performance as in the \textit{TVWS Band Only}. Although, both the \textit{50\% - 50\% Traffic Split} and \textit{Band Switching Per TCP Flow} schemes should have lower performance compared to the \textit{TVWS Band Only} scheme due to not adapting to the unequal channel SNR levels. However, the channel diversity gain in this scenario provides higher throughput to \textit{50\% - 50\% Traffic Split} scheme. The performance of the \textit{Band Switching Per TCP Flow} scheme, where all the packets of a single TCP flow are transmitted through the same band, does not achieve sufficient diversity gain. It is due to the longer use of the same band compared to the the \textit{50\% - 50\% Traffic Split} which makes its throughput performance higher than the \textit{ISM Band Only} but lower than the \textit{TVWS Band only}.  

The \textit{Band Switching Per TCP Flow} scheme, as well as the other single band schemes do not use multiple bands at the same time. Due to that reason, these schemes do not suffer out of order packet delay at the receiver. That leads to a fewer number of packet retransmissions as shown in Figure~\ref{fig:Scenario 1 retrans}. All the aggregation schemes have higher number of packet retransmissions because of exploiting multiple bands simultaneously among which the \textit{Leaky Bucket} provides the best performance as discussed before. 

\subsubsection{High Round Trip Time}
We present a simulation scenario in which we model higher packet delay in the IP cloud module with the values uniformly distributed between 80 to 120 ms. This leads to an end-to-end RTT of 250 ms, which is near to the end-to-end RTT recommended by ITU standard for delay sensitive services like voice or live video streaming~\cite{ITU_standard}. The average MAC latency difference between two bands is reduced to 3 ms. The MAC latency difference between bands is equivalent to 1.2\% of the total RTT.

\begin{figure}
\centering
\includegraphics[width=0.8\linewidth]{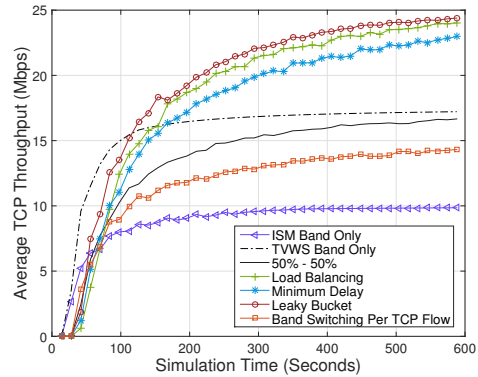}
\caption{Average TCP throughput with long RTT scenario in \textit{Single AP - Single STA} case}
\label{fig:Scenario 2 throughput}
\end{figure}

\begin{figure}
\centering
\includegraphics[width=0.8\linewidth]{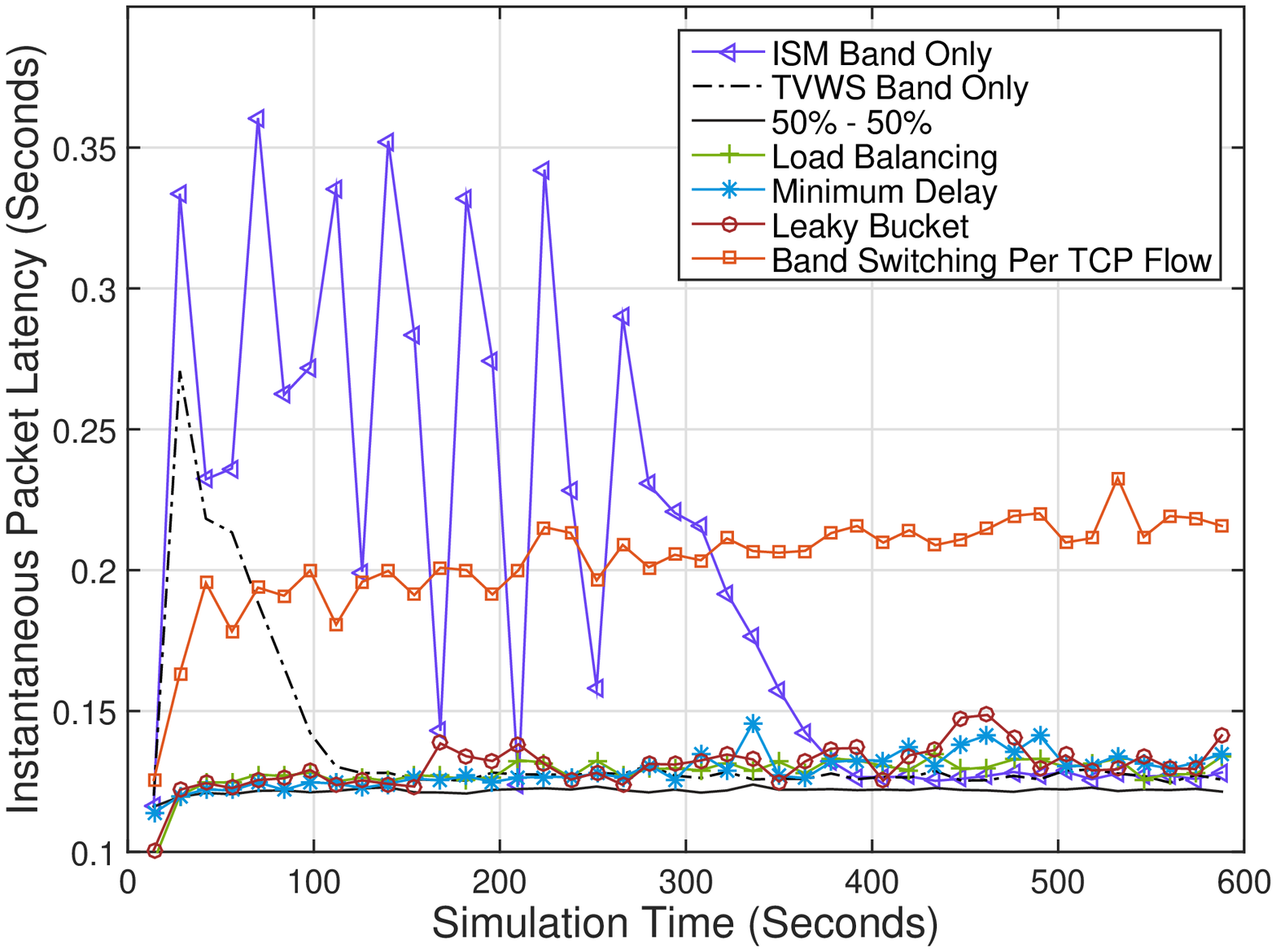}
\caption{Instantaneous packet latency with long RTT scenario in \textit{Single AP - Single STA} case}
\label{fig:Scenario 2 latency}
\end{figure}

\begin{figure}
\centering
\includegraphics[width=0.8\linewidth]{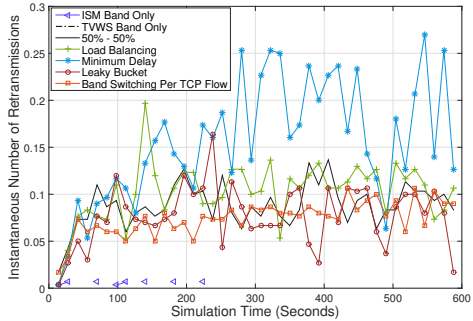}
\caption{Average instantaneous number of retransmissions with long RTT scenario in \textit{Single AP - Single STA} case}
\label{fig:Scenario 2 retrans}
\end{figure}

The simulation results in this scenario are shown in Figures \ref{fig:Scenario 2 throughput}, \ref{fig:Scenario 2 latency} and \ref{fig:Scenario 2 retrans}. The conclusions of the previous case still hold although the performance values are all reduced because of the longer end-to-end RTT. The \textit{TVWS Band Only} scheme always outperforms the \textit{ISM Band Only} scheme. Both schemes suffer high packet latency at the beginning of the simulation but converge later. However, the \textit{ISM Band Only} scheme suffers more initial performance degradation due to the lower SNR condition. In this case the TCP flow control scheme needs more time to absorb the network load.

The \textit{50\% - 50\% Traffic Split} and \textit{Band Switching Per TCP Flow} schemes have lower performance compared to the \textit{TVWS Band Only} scheme since they are unable to adapt to the unequal channel SNR levels. The aggregation schemes have better performance while the \textit{Leaky Bucket} performs the best because of its smart scheduling scheme resulting in lower number of retransmissions as shown in Figure \ref{fig:Scenario 2 retrans}.

% %%%%%%%%%%%    Single AP - Multi Stations       %%%%%%
\subsection{Single AP - Multi Stations}
\begin{figure*}
\centering
\includegraphics[width=4.5in]{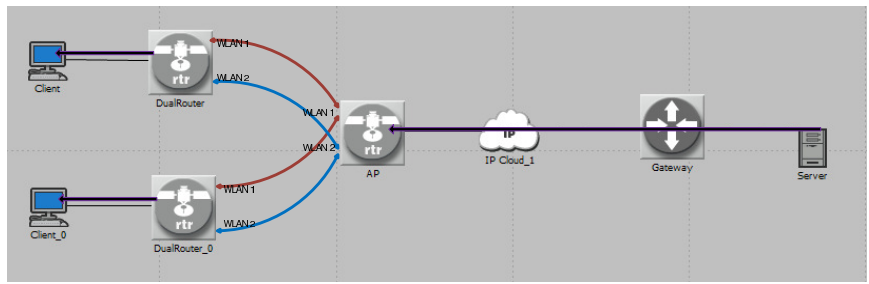}
\caption{Simulation setup for \textit{Single AP - Multi STAs} scenario}
\label{fig:omma_multi_system}
\end{figure*}

\begin{table}[h]
\begin{center}
\begin{tabular}{|c|c|c|}
	\hline
No &  Scheme   &   Offered Load (Mbps) \\
	\hline
1   & 50\% - 50\% Traffic Split & 11.6 \\
2   & Load Balancing & 12 \\
3   & Minimum Delay & 12.4 \\
4  & Leaky Bucket & 12.8 \\
	\hline
\end{tabular}
\end{center}
\caption{\label{tbl:offerload_multi} Offered load for each scheme in \textit{Single AP - Multi STAs} scenario}
\end{table}

\begin{figure}
\centering
\includegraphics[width=0.8\linewidth]{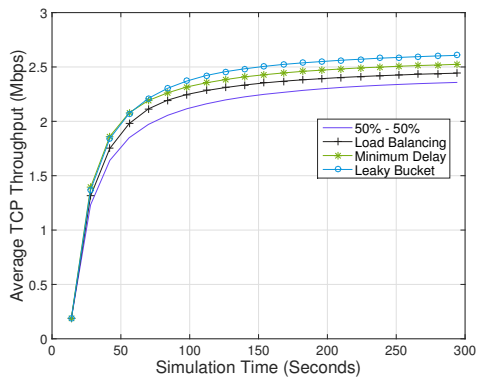}
\caption{Average TCP throughput with small RTT scenario in \textit{Single AP - Multi STAs} case}
\label{fig:Scenario 3 throughput}
\end{figure}

\begin{figure}
\centering
\includegraphics[width=0.8\linewidth]{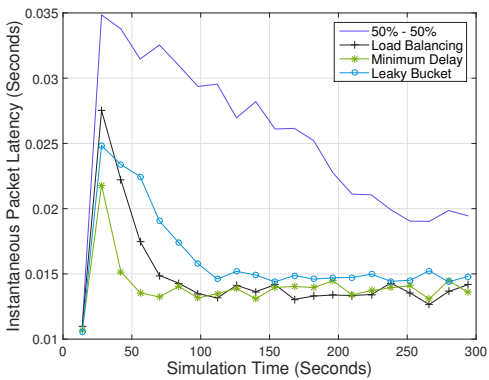}
\caption{Instantaneous packet latency with small RTT scenario in \textit{Single AP - Multi STAs} case}
\label{fig:Scenario 3 latency}
\end{figure}

\begin{figure}
\centering
\includegraphics[width=0.8\linewidth]{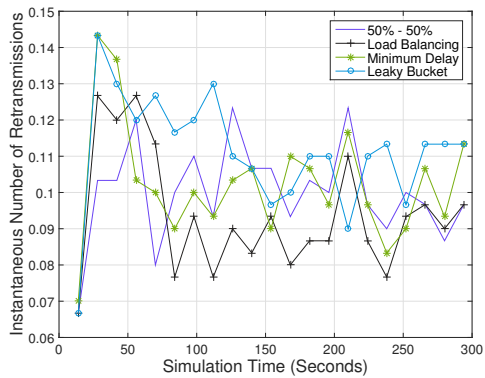}
\caption{Average instantaneous number of retransmissions with small RTT scenario in \textit{Single AP - Multi STAs} case}
\label{fig:Scenario 3 retrans}
\end{figure}

In this section, we present the simulation scenario with a single AP communicating with two STAs simultaneously as shown in Figure \ref{fig:omma_multi_system}. The configurations of AP and STAs are kept similar to the single STA scenario. In this case, at each band, the average vacation time is also measured for each STA's queue. The measurement is taken per packet basis, where the vacation time is calculated for each STA's queue whenever it receives service. This information is continuously fed back to the OMMA layer, which is used in OMMA Leaky Bucket mechanism. We set the interference level of the radio front-end of each band such that the effective average SNR levels at each STA are equal to 10 dB on the ISM band and 20 dB on the TVWS band. We configure STA 1 to operate on both the ISM and TVWS bands, while we configure STA 2 to operate on the ISM band only as a legacy device. STA 1 and STA 2 access the same channels on ISM band, while STA 1 aggregates data on both the ISM and TVWS bands. The IP traffic belongs to the best effort AC (AC\_BE) in this experiment. We implement four scheduling schemes at the OMMA layer: \textit{50\% - 50\% Traffic Split},  \textit{Load Balancing},  \textit{Minimum Delay}, and  \textit{Leaky Bucket}. Similar to the \textit{Single STA} case, the server sends a new file to each STA in every 30s. Similarly the value of the offered load sent to each STA is adjusted for the schemes such that the maximum sustainable load would be achieved. We set the same offered load for each STA and set the packet latency at the IP cloud module to be negligible so that the RTT becomes small. The offered load per STA for each scheduling scheme is shown in Table \ref{tbl:offerload_multi}.

Figure \ref{fig:Scenario 3 throughput} presents the comparison of total throughput sent from the server to both STAs with four different scheduling schemes mentioned above. The \textit{Load Balancing} scheme achieves better throughput than the \textit{50\% - 50\% Traffic Split} since it is a scheduling scheme that adapts with channel quality at the PHY layer. The \textit{Minimum Delay} scheme due to the optimal traffic split scheme at the OMMA layer achieves higher performance than the \textit{Load Balancing} and the \textit{50\% - 50\% Traffic Split} schemes. The \textit{Leaky Bucket} method further improves the performance. The deployment of the unique packet scheduling to reduce the reordering delay at the receiver gives the edge to the \textit{Leaky Bucket} scheme to achieve the best throughput among the tested schemes. In this experiment, the \textit{Leaky Bucket} method's throughput outperforms \textit{Minimum Delay}, \textit{Load Balancing}, and \textit{50\% - 50\% Traffic Split} scheduling schemes by 0.82 Mbps (3.36\%), 1.64 Mbps (6.7\%), and 2.49 Mbps (10.53\%), respectively.

Figure~\ref{fig:Scenario 3 latency} and ~\ref{fig:Scenario 3 retrans} demonstrate the average end-to-end packet latency and number of TCP retransmissions, respectively, of the four scheduling schemes. All four scheduling schemes show high packet latency at the beginning of simulation due to the TCP ramp-up procedure before they all converge. The \textit{Load Balancing} and the \textit{Leaky Bucket} have approximately the similar latency performance, as well as the best among all four schemes. 

In terms of number of TCP retransmissions, all four schemes have approximately the similar percentage of TCP packet retransmissions of around 10\% for the given offered network loads as shown in Table \ref{tbl:offerload_multi}. Among all schemes, \textit{Load Balancing} has the lowest number of packet retransmissions, where the maximum average performance gap between the \textit{Leaky Bucket} and the \textit{Load Balancing} is near to 2\%. This is an artifact of the 7\% higher offered load in the \textit{Leaky Bucket} as given in Table \ref{tbl:offerload_multi}. Under the same offered load as in the case of previous section, the \textit{Leaky Bucket} scheme had the best performance in terms of latency as well as the number of retransmissions. Therefore, in this case, we can conclude that \textit{Leaky Bucket} maintains the highest offered load with a small penalty in the number of retransmissions, however, the large gain in throughput still makes it the best choice among all the schemes. 

%%%%%%%%%%%%%%%%%%%%%%%%%%%%%%%%%%%%%%%%%%%
% %%%%%%%%%%%   Conclusion      %%%%%%
\section{Conclusion}\label{conclusion}
We presented an analytical framework to derive the optimal packet scheduling strategy over multiple bands for WLAN systems. For such system an optimal packet distribution scheme minimizing the average end-to-end packet latency is designed. We further proposed a per-packet scheduling algorithm, called OMMA Leaky Bucket. It not only distributes the packets over multi bands using the derived optimal distribution, but also minimizes re-sequencing delay at the receiver. We also described the OMMA system architecture. It includes a functional design, a solution for discovery and association procedures between multi-RAT devices, and a dynamic RAT update management entity. Finally, we provided a set of simulations including both single AP - single STA and single AP - multi STAs scenarios while comparing the proposed OMMA Leaky Bucket approach to various alternatives. The proposed OMMA Leaky Bucket  scheme outperforms all other schemes in terms of throughput, packet latency, and number of retransmission in all the cases that we have considered. 

For future work, we will include the analysis of our system for a more realistic traffic model with non-Markov properties. We will also include the performance evaluation under such conditions with different QoS requirements. 
%%%%%%%%%%%%%%%%%%%%%%%%%%%%%%%%%%%%%%%%%%%
% %%%%%%%%%%%   References      %%%%%%
\bibliographystyle{IEEEtran}
\bibliography{omma}

% Generated by IEEEtran.bst, version: 1.13 (2008/09/30)
\begin{thebibliography}{10}
\providecommand{\url}[1]{#1}
\csname url@samestyle\endcsname
\providecommand{\newblock}{\relax}
\providecommand{\bibinfo}[2]{#2}
\providecommand{\BIBentrySTDinterwordspacing}{\spaceskip=0pt\relax}
\providecommand{\BIBentryALTinterwordstretchfactor}{4}
\providecommand{\BIBentryALTinterwordspacing}{\spaceskip=\fontdimen2\font plus
\BIBentryALTinterwordstretchfactor\fontdimen3\font minus
  \fontdimen4\font\relax}
\providecommand{\BIBforeignlanguage}[2]{{%
\expandafter\ifx\csname l@#1\endcsname\relax
\typeout{** WARNING: IEEEtran.bst: No hyphenation pattern has been}%
\typeout{** loaded for the language `#1'. Using the pattern for}%
\typeout{** the default language instead.}%
\else
\language=\csname l@#1\endcsname
\fi
#2}}
\providecommand{\BIBdecl}{\relax}
\BIBdecl

\bibitem{OMMA1}
T.~Elkourdi, A.~Chincholi, T.~Le, and A.~Demir, ``Cross-layer optimization for
  opportunistic multi-{MAC} aggregation,'' in \emph{IEEE Vehicular Technology
  Conference (VTC Spring)}, June 2013, pp. 1--5.

\bibitem{Intel_multiRAT}
N.~Himayat, S.-P. Yeh, A.~Panah, S.~Talwar, M.~Gerasimenko, S.~Andreev, and
  Y.~Koucheryavy, ``Multi-radio heterogeneous networks: architectures and
  performance,'' in \emph{Computing, Networking and Communications (ICNC), 2014
  International Conference on}, Feb 2014, pp. 252--258.

\bibitem{ZhangVMH10}
D.~Zhang, P.~K. Vitthaladevuni, B.~Mohanty, and J.~Hou, ``Performance analysis
  of dual-carrier {HSDPA},'' in \emph{IEEE Vehicular Technology Conference (VTC
  Spring)}, 2010, pp. 1--5.

\bibitem{KoudouridisYK09}
G.~P. Koudouridis, A.~Yaver, and M.~U. Khattak, ``Performance evaluation of
  multi-radio transmission diversity for {TCP} flows,'' in \emph{Vehicular
  Technology Conference (VTC Spring)}, 2009.

\bibitem{GLL2015}
G.~P. Koudouridis, H.~Lundqvist, H.~R. Karimi, and G.~Karlsson, ``A
  quantitative analysis of the throughput gains and the energy efficiency of
  multi-radio transmission diversity in dense access networks,''
  \emph{Telecommun. Syst.}, vol.~59, no.~1, pp. 145--168, May 2015.

\bibitem{Netgear}
\BIBentryALTinterwordspacing
NETGEAR, ``Why choose simultaneous dual band?'' 2011. [Online]. Available:
  \url{http://www.netgear.com/landing/dual-band.aspx}
\BIBentrySTDinterwordspacing

\bibitem{Han2006}
H.~Han, S.~Shakkottai, C.~V. Hollot, R.~Srikant, and D.~Towsley, ``Multi-path
  {TCP}: a joint congestion control and routing scheme to exploit path
  diversity in the internet,'' \emph{IEEE/ACM Transactions on Networking},
  vol.~14, no.~6, pp. 1260--1271, Dec. 2006.

\bibitem{Kelly2005}
F.~Kelly and T.~Voice, ``Stability of end-to-end algorithms for joint routing
  and rate control,'' \emph{ACM SIGCOMM Computer Communication Review},
  vol.~35, no.~2, pp. 5--12, Apr. 2005.

\bibitem{FRHB11}
A.~Ford, C.~Raiciu, M.~Handley, and O.~Bonaventure, ``{TCP} extensions for
  multipath operation with multiple addresses,'' RFC 6824, Jan. 2013.

\bibitem{Key_multipathrouting}
P.~Key, L.~Massoulie, and D.~Towsley, ``Multipath routing, congestion control
  and dynamic load balancing,'' in \emph{IEEE International Conference on
  Acoustics, Speech and Signal Processing (ICASSP)}, vol.~4, April 2007, pp.
  IV--1341--IV--1344.

\bibitem{Generic_link_layer_1}
G.~Koudouridis, R.~Agüero, E.~Alexandri, J.~Choque, K.~Dimou, H.~Arimi,
  H.~Lederer, J.~Sachs, and R.~Sigle, ``Generic link layer functionality for
  multi-radio access networks,'' in \emph{IST Mobile and Wireless
  Communications Summit}, 2005.

\bibitem{Generic_link_layer_2}
K.~Dimou, R.~Agero, M.~Bortnik, R.~Karimi, G.~Koudouridis, S.~Kaminski,
  H.~Lederer, and J.~Sachs, ``Generic link layer: a solution for multi-radio
  transmission diversity in communication networks beyond {3G},'' in \emph{IEEE
  Vehicular Technology Conference (VTC Fall)}, 2005.

\bibitem{HetNet}
\BIBentryALTinterwordspacing
(2015) {IEEE} 802.1 {OmniRAN} task group. [Online]. Available:
  \url{https://mentor.ieee.org/omniran/bp/StartPage}
\BIBentrySTDinterwordspacing

\bibitem{Zhang2015}
H.~Zhang, C.~Jiang, N.~C. Beaulieu, X.~Chu, X.~Wang, and T.~Q.~S. Quek,
  ``Resource allocation for cognitive small cell networks: A cooperative
  bargaining game theoretic approach,'' \emph{IEEE Transactions on Wireless
  Communications}, vol.~14, no.~6, pp. 3481--3493, June 2015.

\bibitem{Zheng2016}
Q.~Zheng, K.~Zheng, H.~Zhang, and V.~C.~M. Leung, ``Delay-optimal virtualized
  radio resource scheduling in software-defined vehicular networks via
  stochastic learning,'' \emph{IEEE Transactions on Vehicular Technology},
  vol.~65, no.~10, pp. 7857--7867, Oct 2016.

\bibitem{Zhang2014}
H.~Zhang, C.~Jiang, N.~C. Beaulieu, X.~Chu, X.~Wen, and M.~Tao, ``Resource
  allocation in spectrum-sharing ofdma femtocells with heterogeneous
  services,'' \emph{IEEE Transactions on Communications}, vol.~62, no.~7, pp.
  2366--2377, July 2014.

\bibitem{Radio_Resource_Switching}
X.-D. Trinh, G.~Jo, J.~Lee, J.-H. Na, W.~Park, and H.-S. Cho, ``A
  radio-resource switching scheme in aggregated radio access network,'' in
  \emph{The Seventh International Conference on Digital Telecommunications},
  2012.

\bibitem{choi2010joint}
Y.~Choi, H.~Kim, S.-w. Han, and Y.~Han, ``Joint resource allocation for
  parallel multi-radio access in heterogeneous wireless networks,''
  \emph{Wireless Communications, IEEE Transactions on}, vol.~9, no.~11, pp.
  3324--3329, 2010.

\bibitem{chen2009opportunistic}
F.~Chen, H.~Zhai, and Y.~Fang, ``An opportunistic multiradio {MAC} protocol in
  multirate wireless ad hoc networks,'' \emph{Wireless Communications, IEEE
  Transactions on}, vol.~8, no.~5, pp. 2642--2651, 2009.

\bibitem{cui2009novel}
Y.~Cui, Y.~Xu, X.~Sha, R.~Xu, and Z.~Ding, ``A novel multi-radio packet
  scheduling algorithm for real-time traffic on generic link layer,'' in
  \emph{Communications, 2009. APCC 2009. 15th Asia-Pacific Conference
  on}.\hskip 1em plus 0.5em minus 0.4em\relax IEEE, 2009, pp. 122--125.

\bibitem{Koudouridis2016}
G.~P. Koudouridis, P.~Soldati, and G.~Karlsson, ``Multiple connectivity and
  spectrum access utilisation in heterogeneous small cell networks,''
  \emph{International Journal of Wireless Information Networks}, vol.~23,
  no.~1, pp. 1--18, 2016.

\bibitem{Zhou2016}
Y.~Zhou, J.~Chen, and Y.~Kuo, ``Fairness resource allocation for parallel
  multi-radio access in cognitive multi-cell,'' \emph{Wireless Personal
  Communications}, vol.~88, no.~3, pp. 587--602, 2016.

\bibitem{WuVKHC11}
Y.~Wu, H.~Viswanathan, T.~E. Klein, M.~Haner, and A.~R. Calderbank, ``Capacity
  optimization in networks with heterogeneous radio access technologies,'' in
  \emph{IEEE Global Telecommunications Conference (GLOBECOM)}, 2011, pp. 1--5.

\bibitem{KonIHHIH12}
Y.~Kon, M.~Ito, N.~Hassel, M.~Hasegawa, K.~Ishizu, and H.~Harada, ``Autonomous
  parameter optimization of a heterogeneous wireless network aggregation system
  using machine learning algorithms,'' in \emph{IEEE Consumer Communications
  and Networking Conference (CCNC)}, 2012, pp. 894--898.

\bibitem{Concurrent_Bandwidth_Aggregation}
D.~Krishnaswamy, D.~Zhang, S.~Soliman, B.~Mohanty, D.~Cavendish, W.~Ge, and
  S.~Eravelli, ``Concurrent bandwidth aggregation over wireless networks,'' in
  \emph{IEEE International Conference on Computing, Communications, and
  Networking (ICNC)}, 2012.

\bibitem{Survey2012}
A.~L. Ramaboli, O.~E. Falowo, and A.~H. Chan, ``Bandwidth aggregation in
  heterogeneous wireless networks: A survey of current approaches and issues,''
  \emph{Journal of Network and Computer Applications}, vol.~35, no.~6, pp.
  1674--1690, 2012.

\bibitem{Broadcom}
\BIBentryALTinterwordspacing
CNET, ``Broadcom's new real dual-band wi-fi chip speeds things up,'' 2015.
  [Online]. Available:
  \url{http://www.cnet.com/news/broadcoms-new-real-dual-band-wi-fi-chip-speeds-things-up}
\BIBentrySTDinterwordspacing

\bibitem{IEEE802.11n}
\emph{Wireless {LAN} Medium Access Control {(MAC)} and Physical Layer {(PHY)}
  Specification Amendment 5: Enhancements for Higher Throughput}, IEEE Std.
  802.11n, 2009.

\bibitem{IEEE802.11ac}
\emph{Wireless {LAN} Medium Access Control {(MAC)} and Physical Layer {(PHY)}
  Specification Amendment 4: Enhancements for Very High Throughput}, IEEE Std.
  802.11ac, 2013.

\bibitem{IEEE802.11e}
\emph{Wireless {LAN} Medium Access Control {(MAC)} and Physical Layer {(PHY)}
  Specification Amendment 8: Medium Access Control (MAC) Quality of Service
  Enhancements}, IEEE Std. 802.11e, 2005.

\bibitem{data_networks}
D.~P. Bertsekas and R.~Gallager, \emph{Data networks (2. ed.)}.\hskip 1em plus
  0.5em minus 0.4em\relax Prentice Hall, 1992.

\bibitem{next_gen_wlan}
P.~Eldad and S.~Robert, \emph{Throughput, Robustness, and Reliability in
  802.11n}.\hskip 1em plus 0.5em minus 0.4em\relax Cambridge University Press.,
  2008.

\bibitem{opt_theo}
D.~A. Pierre, \emph{Optimization Theory with Applications}.\hskip 1em plus
  0.5em minus 0.4em\relax Dover Publications, 1986.

\bibitem{QoS_reorder_1}
Z.~Yan, M.~Veeraraghavan, C.~Tracy, and C.~Guok, ``On how to provision quality
  of service {(QoS)} for large dataset transfers,'' in \emph{Proceedings of the
  Sixth International Conference on Communication Theory, Reliability, and
  Quality of Service (CTRQ)}.\hskip 1em plus 0.5em minus 0.4em\relax IARIA,
  2013, pp. 21--26.

\bibitem{QoS_reorder_2}
X.~Zhou and P.~Van~Mieghem, ``Reordering of ip packets in internet,'' in
  \emph{Passive and Active Network Measurement}.\hskip 1em plus 0.5em minus
  0.4em\relax Springer, 2004, pp. 237--246.

\bibitem{vlavianos2008}
A.~Vlavianos, L.~K. Law, I.~Broustis, S.~V. Krishnamurthy, and M.~Faloutsos,
  ``Assessing link quality in {IEEE} 802.11 wireless networks: which is the
  right metric?'' in \emph{Personal, Indoor and Mobile Radio Communications,
  2008. PIMRC 2008. IEEE 19th International Symposium on}.\hskip 1em plus 0.5em
  minus 0.4em\relax IEEE, 2008, pp. 1--6.

\bibitem{schwartz1996broadband}
M.~Schwartz, \emph{Broadband integrated networks}.\hskip 1em plus 0.5em minus
  0.4em\relax Prentice Hall PTR New Jersey, 1996, vol.~19.

\bibitem{ITU_standard}
\emph{End-user multimedia QoS categories}, ITU Std. G.1010, 2001.

\end{thebibliography}
 \vspace{-10mm}
\begin{IEEEbiography}[{\includegraphics[width=1in,height=1.25in,clip,keepaspectratio]{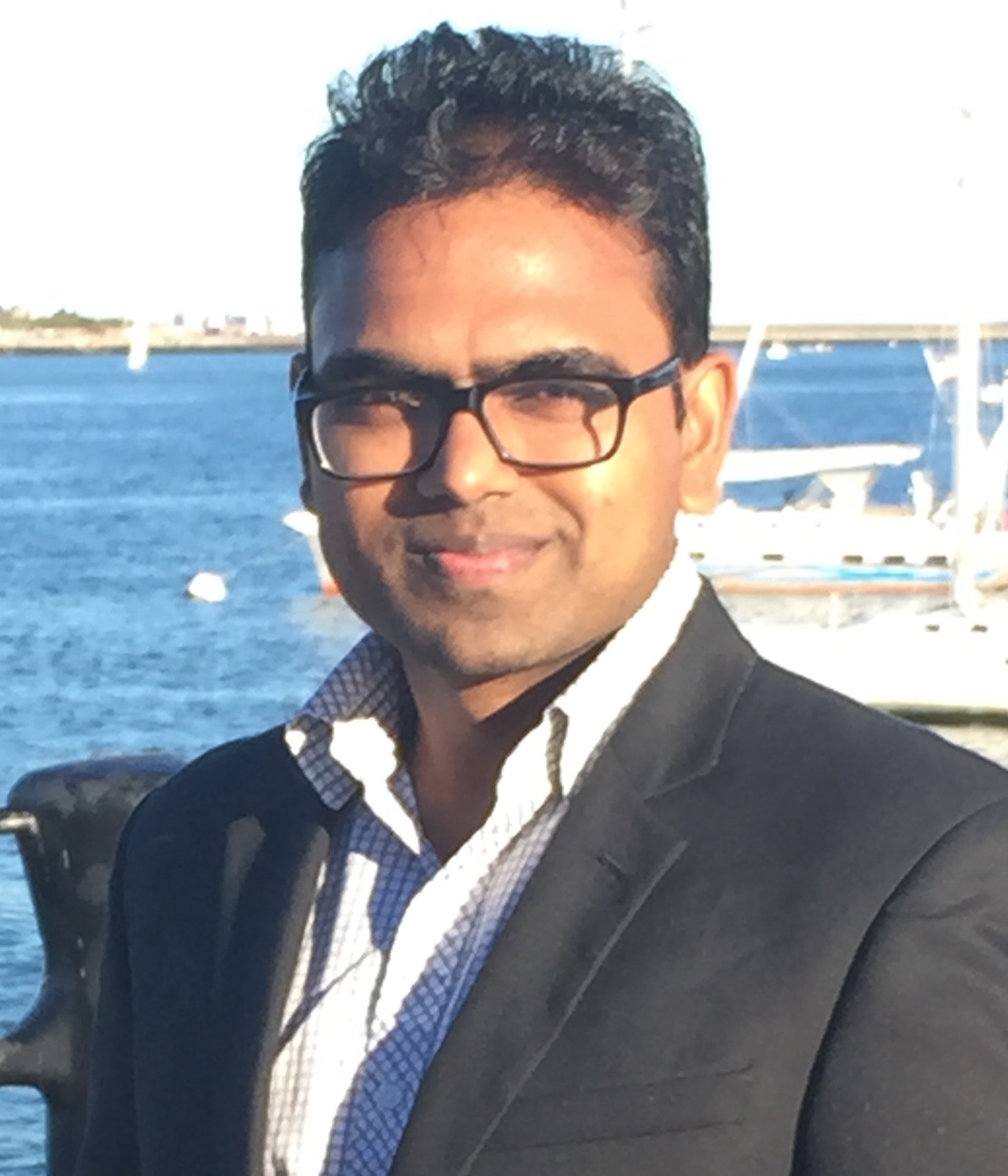}}]{Sanjay Goyal}
received his Ph.D. and M.S. degrees in electrical engineering from NYU Tandon School of Engineering, New York, in 2016 and 2012, respectively. He received his B.Tech. degree in communication and computer engineering from the LNM Institute of Information Technology, India, in 2009. Currently, he is a post graduate researcher in InterDigital Communications. He was awarded, along with Carlo Galiotto, Nicola Marchetti, and Shivendra Panwar, the Best Paper Award in IEEE ICC 2016. His research interests are in designing and analyzing wireless network protocols with an emphasis on cross-layer optimization, especially with the PHY and MAC layers.
\end{IEEEbiography}  \vspace{-10mm}
\begin{IEEEbiography}[{\includegraphics[width=1.1in,height=1.25in,clip,keepaspectratio]{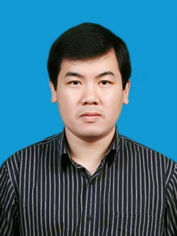}}]{Tan Ba Le}
received his Ph.D. degree Electrical Engineering from Polytechnic Institute of New York University in 2011. He received his Engineering and M.S. degrees in Electronics and Telecommunications from Hanoi University of Science and Technology, Viet Nam, in 2000 and 2003 respectively. He was part of a team developing technologies on spectrum sharing from 2011 to 2014 as a Senior Engineer at InterDigital Comm. Corp. He is currently the Head of Technology Department at Viettel Group in Vietnam.
\end{IEEEbiography} \vspace{-10mm}
\begin{IEEEbiography}[{\includegraphics[width=1in,height=1.25in,clip,keepaspectratio]{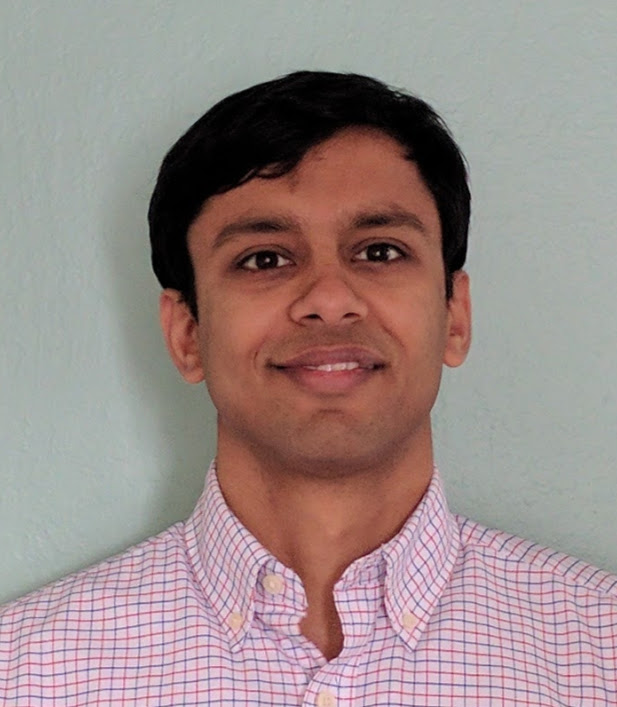}}]{Amith Chincholi}
received his M.S. degree in Electrical and Computer Engineering from Rutgers, The State University of New Jersey, New Brunswick and his B.E. degree from Osmania University College of Engineering (Autonomous), India. He has more than 12 years of experience in wireless system design involving cutting-edge modem development for 3G HSPA, 4G LTE/LTE-A, dual-SIM concurrent RAT and NB-IoT, and, bleeding edge research and technology generation for next generation wireless systems while working at InterDigital Communications and Qualcomm Incorporated. He also represents Qualcomm at the 3GPP RAN4 cellular standards meetings and is involved in standardization of various advanced technology features which includes writing many technical contributions, working with other company delegates to generate consensus on the feature definition and drafting the 3GPP standards? specification to define the feature. He is currently involved in NB-IoT standardization and also leads NB-IoT modem algorithm development, software and firmware implementation.
\end{IEEEbiography} \vspace{-10mm}
\begin{IEEEbiography}[{\includegraphics[width=1.1in,height=1.25in,clip,keepaspectratio]{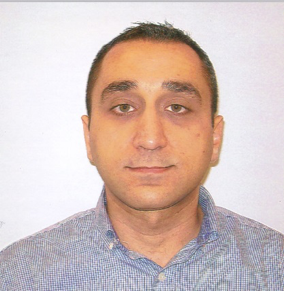}}]{Tariq Elkourdi}
received the Ph.D. degree in Electrical Engineering and the M. Sc. degree in Telecommunications Engineering from New Jersey Institute of Technology in 2012 and 2008, respectively, and the B.Sc. degree in Communications and Electronics Engineering from the Applied Science University, Jordan, in 2006. Currently, he is an Adjunct Professor of Electrical and Computer Engineering at New Jersey Institute of Technology and is affiliated with Nokia Mobile Networks. His research interests concern wireless communications, optimization and emerging wireless technologies.
\end{IEEEbiography}  \vspace{-10mm}
\begin{IEEEbiography}[{\includegraphics[width=1.1in,height=1.25in,clip,keepaspectratio]{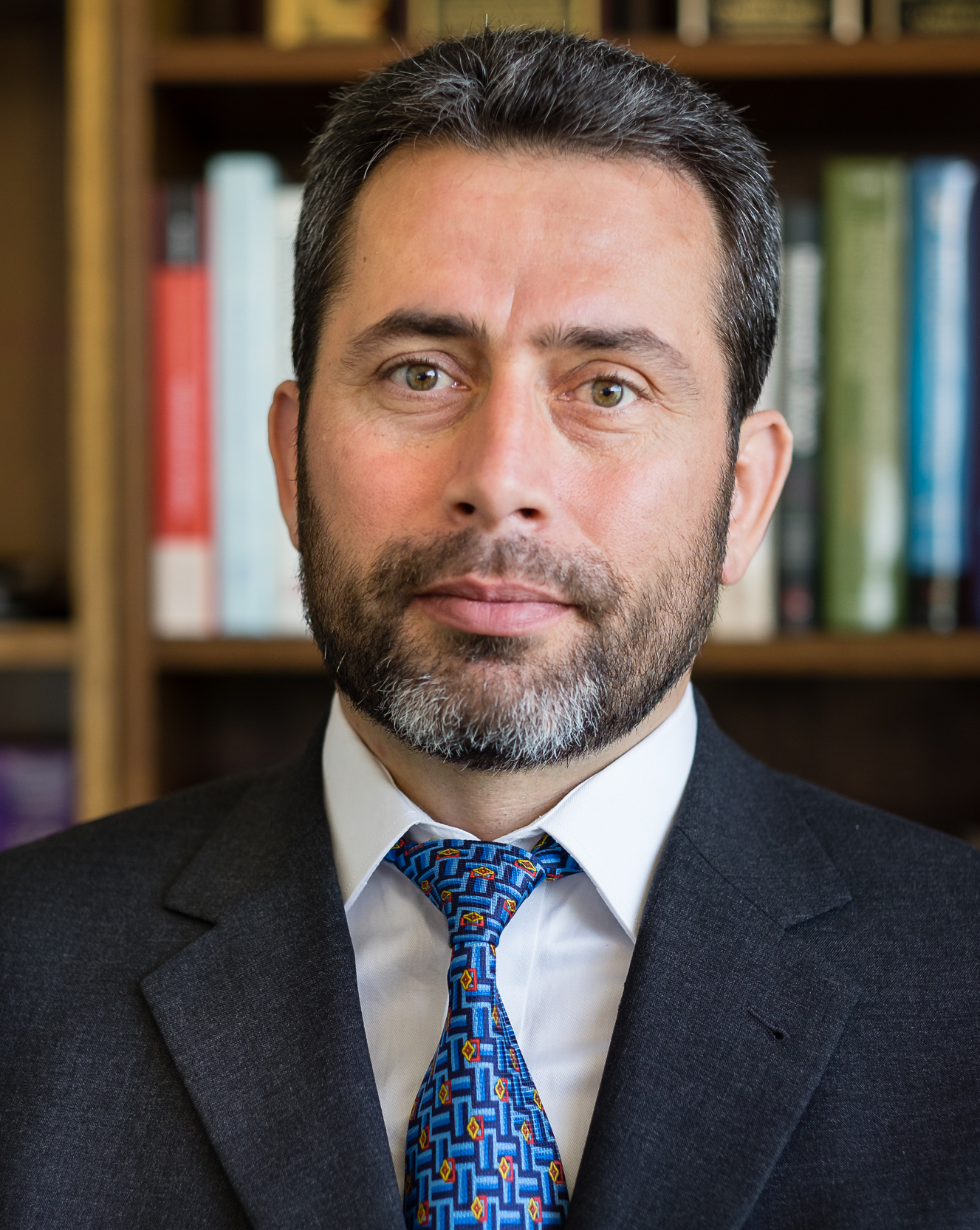}}]{Alpaslan Demir}
is a Principal Engineer in the Future Wireless Business Group at InterDigital Comm. Corp. He is part of a team working on Next Generation Access and currently focusing on activities related to 5G and beyond. He has been serving the wireless communications industry for more than twenty years with a unique combination of experiences relevant to MAC, PHY, and RF design. Notably, he is a prolific inventor with 59 granted and numerous pending patent applications to date. He is an IEEE member and holds a Ph.D. degree in Communications from Polytechnic Institute of NYU.

\end{IEEEbiography}

\end{document}